\documentclass[twocolumn,aps,prl,superscriptaddress]{revtex4-2}

\usepackage{graphicx}  
\usepackage{bm}        
\usepackage{amssymb}   
\usepackage{xcolor}
\usepackage{amsmath}
\usepackage{braket}
\definecolor{linkColor}{RGB}{0,80,150}
\usepackage{hyperref}
\usepackage{algpseudocode}
\usepackage{algorithm}
\usepackage{wrapfig}   
 \usepackage{orcidlink}  

\algrenewcommand\algorithmicrequire{\textbf{Input:}}
\algrenewcommand\algorithmicensure{\textbf{Output:}}
\usepackage{varwidth}

\usepackage[normalem]{ulem}
\usepackage{soul}

\hypersetup{
    colorlinks=true,
    allcolors=linkColor,
    pdfborder={0 0 0},
    pdfencoding = auto
}

\makeatletter
\newcommand*{\rom}[1]{\expandafter\@slowromancap\romannumeral #1@}
\makeatother
 
\begin{document}

\title{Non-Equilibrium Origin of Native Ring Anisotropy in Amorphous Systems}

\author{Zihang Wang\,\orcidlink{0000-0003-4482-117X}}

\email{zihangwang@ucsb.edu}

\affiliation{Department of Physics, University of California Santa Barbara, Santa Barbara, California 93106, USA}
 
 \author{Dirk Bouwmeester\,\orcidlink{0000-0002-2118-6532}}

\affiliation{Department of Physics, University of California Santa Barbara, Santa Barbara, California 93106, USA}
 \affiliation{Huygens-Kamerlingh Onnes Laboratory, Leiden University, P.O. Box 9504, 2300 RA Leiden, Netherlands}

\begin{abstract}
Native ring structures within amorphous networks play a critical role in determining structural and optical properties, in part due to their ability to host dopants such as rare earth ions in silicate systems. In this work, we demonstrate that the universal features of structural anisotropy in amorphous networks can be efficiently simulated using a model based on stochastically deformed, edge-sharing N-member native ring structures (N-NRS). This model isolates and characterizes the structural anisotropy generated during the annealing–quenching process that is independent of any constituent-specific interactions. We refer to this computational framework as \textit{Indistinguishable Simulated Folding} (ISF), a stochastic process that mimics a simulated annealing–quenching procedure. Formulated as a Markov process, ISF is governed by two physically meaningful parameters: the number of Markov steps, representing the mean duration of each ring-folding event, and the stochastic deformation magnitude, which quantifies thermally induced structural changes per event. Furthermore, we show that the logarithm of any positive-valued anisotropy measure generated by ISF is a skewed random variable, reflecting the growing entropy production rate during the Markov evolution. ISF provides both a conceptual framework for understanding the universal stochastic origin of structural anisotropy in amorphous networks and a practical tool for simulating constituent-independent features, without requiring full-scale molecular dynamics simulations.

\end{abstract}

\maketitle

\definecolor{lightblue}{RGB}{200,220,255}
\setlength{\fboxsep}{0.03\linewidth}
\noindent\fcolorbox{white}{white}{\parbox{0.94\linewidth}{%
}}
\vspace{0.5em}

\section{\rom{1}. Introduction }
 
Native amorphous materials, such as amorphous silicon nitride (SiN) and silica (SiO), lack translational symmetry and therefore exhibit only short range order. They are synthesized through various non-equilibrium processes that inhibit crystallization, such as rapid quenching. An important feature of amorphous systems is the presence of native ring structures, where oxygen and silicon atoms alternate to form closed loops within the amorphous network \cite{Valladarest1999,ELHAYEK2020110425,GUTTMAN1990145}. Native ring structures within an amorphous solid tend to cluster together, forming larger aggregates, and this ring agglomeration process might be responsible for the formation of the amorphous network \cite{KERNER19959}. In addition, native ring structures and their anisotropy play a crucial role in shaping the physical properties of amorphous systems, influencing aspects such as mechanical strength \cite{PhysRevB.107.144203}. Furthermore, the ring structures can be experimentally observed and quantified via spectral analysis, such as defect lines at D1 (\(495 \ \mathrm{cm}^{-1}\)) and D2 (\(606 \ \mathrm{cm}^{-1}\)) \cite{PhysRevLett.80.5145,SITARZ1999281}, or via pair distribution functions constructed through x-ray and neutron diffraction experiments \cite{PhysRevLett.69.1387,GRIMLEY199049,Wang2015,Grigoriev2015,DOVE2022100037,PhysRevB.54.15808,BISWAS20181}.

Native ring structures can serve as hosts for low density interstitial dopants without significantly altering the amorphous network, enabling various applications. For example, by doping a small amount of interstitial aluminum, the amorphous host exhibits much lower oxidation and corrosion rates \cite{Wang2006,https://doi.org/10.1002/adem.200400010}. In addition to the material enhancement, numerical studies hint that native ring structures could also serve as interstitial cages for rare earth ion emitters (REIs) \cite{sun_geometric_2019,LIU2022112388,LIU2020110030,Fuji}, and have been used to generate a wide range of optical applications such as erbium doped fiber amplifiers that enhance optical signals in long haul telecommunications networks \cite{10.1063/5.0165339}, and fiber lasers with REIs that demonstrate high electrical to optical conversion efficiency and reliable spectral variability \cite{zhou2022towards}.

Furthermore, interstitial REIs are promising candidates for various quantum applications since the localized \( f \) orbitals in REIs do not strongly hybridize with the orbitals of neighboring atoms, resulting in narrow homogeneous emission linewidths \cite{serrano2022ultra,doi:10.1021/acs.nanolett.9b03831}. In particular, the local anisotropic field environment created by native ring structures lifts \( 4f \) degeneracy of the embedding REIs, enabling long coherence times for \( 4f \) to \( 4f \) optical transitions, making them suitable for quantum storage and quantum communication when coupled to an optical cavity \cite{ding2016multidimensional}.

Although there has been recent progress in quantifying native ring anisotropy via principal eigen analysis \cite{Shiga2023}, the generating mechanism of native ring anisotropy remains unclear \cite{LEWIS2022121383}.{A key question is how to quantify the relative contributions of stochastic inelastic deformation accumulation during annealing and particle specific contributions to the overall structural anisotropy. Understanding the universal properties of amorphous materials is important for exploring novel applications.} For example, due to the short range order in amorphous hosts, native ring structures with similar anisotropy could create nearly identical local field environments for interstitial REIs, which could have advantages for reducing inhomogeneous and homogeneous linewidths of rare earth ions doped in amorphous systems. This unique property allows amorphous hosts to function not only as gain media for high efficiency lasers but also potentially as scalable quantum light-matter interfaces.

In this work, we demonstrate that the structural anisotropy statistics of an amorphous network can be recovered through an ensemble of edge sharing N-member native ring structures (N-NRS) within the network. We propose the concept of indistinguishable simulated folding (ISF), a minimal Markov chain model that generates N-NRS with statistical properties closely resembling the{constituent independent contribution} of N-NRS sampled from molecular dynamics simulations, without the need to incorporate atomic interactions. Numerical and mathematical evidence suggests that, under certain conditions, the logarithm of any positively defined structural anisotropy measure generated by ISF is a skewed random variable that can be decomposed into diffusion-like and drift-like stochastic contributions, which are directly related to the change of entropy production rate. ISF is a stochastic process governed by two physical parameters: the number of Markov steps and the stochastic deformation magnitude, which can be intuitively understood as the mean duration of each scattering event and the thermally induced structural deformations, respectively. These quantities directly contribute to the total quenching rate.{We investigate two idealized scenarios, the adiabatic and rapid regimes. In both cases, thermal fluctuations solely induce structural deformations.} These contributions are critical for understanding and designing next generation amorphous materials with extended functionalities.

The paper is organized as follows: In Section~\rom{2}, we emphasize the importance of various N-member native ring structures (N-NRS) in capturing the structural anisotropy of the hosting amorphous network. In Section~\rom{3}, we introduce various anisotropy measures, such as nearest neighbor distance, roundness, roughness, and deformation distance. In Section~\rom{4},~\rom{5},~\rom{6}, we present the concept of indistinguishable simulated folding (ISF), which generates N-NRS ensembles that reveal a close structural agreement with the{constituent-independent contribution} of the MD generated N-NRS ensembles. In Section~\rom{7},~\rom{8},~\rom{9}, we show several important properties that fundamentally connect ISF to stochastic state evolution.

\begin{figure*}[t!]
    \centering
    \includegraphics{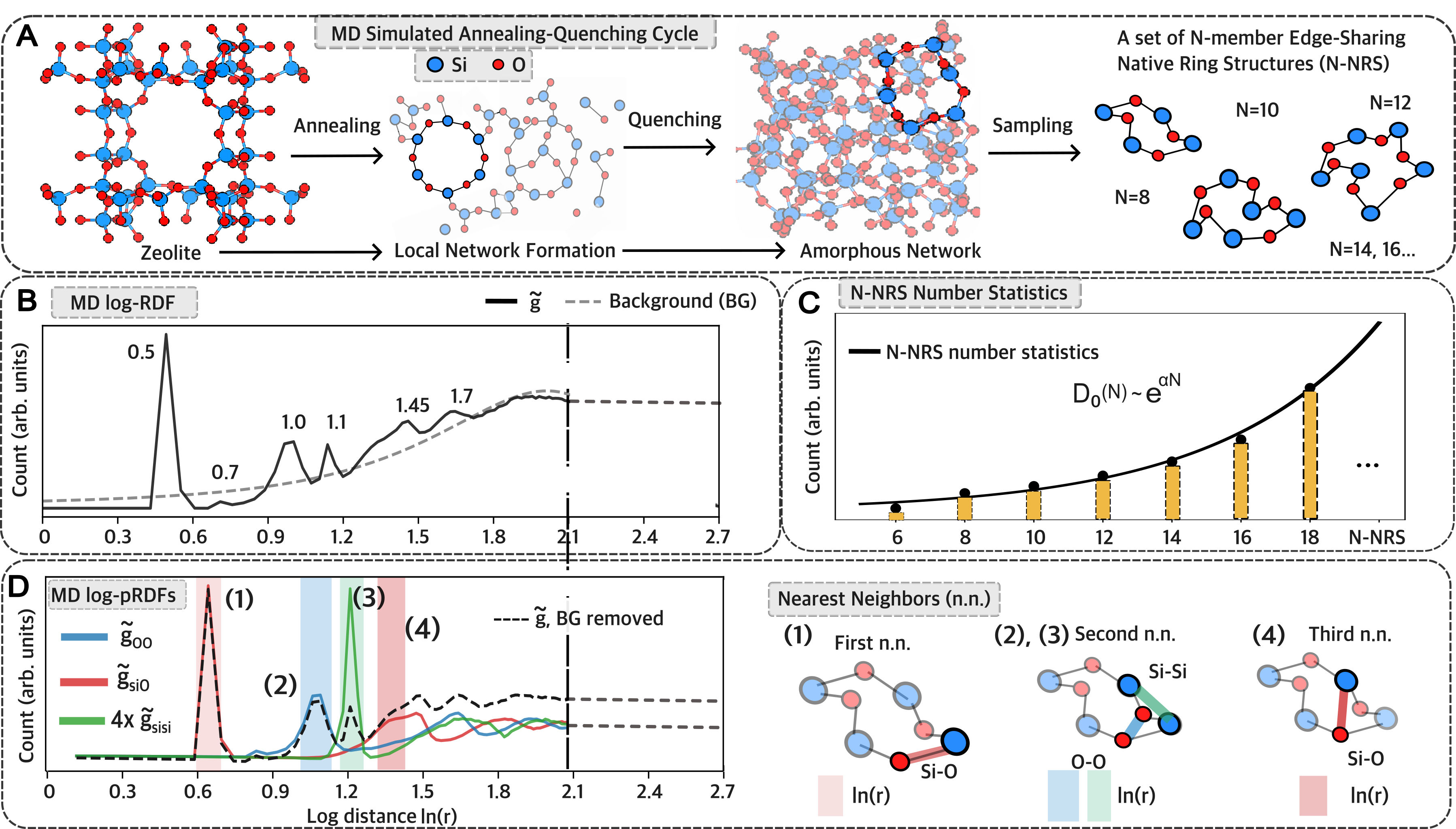}
    \caption{\textbf{A} The simulated annealing quenching cycle and the formation of the amorphous network via molecular dynamics (MD) with the Tersoff pair potential. A prepared silica zeolite structure is initially set up to perform the annealing. The quenching process redistributes atomic arrangements in the network, leading to an amorphous network at the end. For each N, edge sharing N-member native ring structures (N-NRS) are iteratively sampled from the resulting network (fourth schematic) in Eq.~\ref{ensemble}. \textbf{B} The black curve is the log radial distribution function (log-RDF, defined in Eq.~\ref{RDF}) of the above MD simulated amorphous network, and the dashed line is the fitted background resulting from the finite size effect. The geometric distances at log-RDF peaks are denoted. \textbf{C} The number statistics of nonprimitive N-NRSs from the MD simulated amorphous network, represented by the orange bars via the unrestricted counting method. As discussed in the text, the N-NRS number statistics exhibit an exponential divergence with respect to N (the black curve). \textbf{D} The log partial radial distribution function (log-pRDF) captures atom specific pair correlations. In amorphous silica, there are three correlation pairs: silicon-oxygen (Si-O, red curve), oxygen-oxygen (O-O, blue curve), and silicon-silicon (Si-Si, green curve). Four primary regions are indexed (1-4) and shaded by the corresponding colors, representing different types of nearest neighbor correlations, as demonstrated via schematics on the right side. The background is subtracted to ensure \( \lim_{r \to \infty} g[\ln(r)] = 1 \), limited by the vertical dashed line across \textbf{B,D}. }
    \label{fig1}
\end{figure*}

\section{\rom{2}. Edge-sharing Native Ring Structures in the amorphous network}
Local anisotropy in an amorphous network is a critical factor in understanding dopant induced structural variations during an annealing quenching cycle. The radial distribution function (RDF) encodes essential information about structural anisotropy through density variations relative to a reference atom, reflecting the local structural arrangement of the amorphous network. It is defined as:
\begin{equation}
    g(r) \sim \sum_{j} \sum_{i \neq j} \delta(r - |\mathbf{x}_{i} - \mathbf{x}_{j}|),
\end{equation}
where \( r \) is a positively defined variable representing the pairwise distances between any two atoms, and \( \lim_{r \to \infty} g(r) = 1 \) due to spatial homogeneity. The RDF is closely related to the structure factor via a Fourier transform and is experimentally accessible \cite{10.1093/jmicro/dfaa075,doi:10.1021/acs.chemrev.1c00237}.

To investigate the edge sharing NRSs in the amorphous network, as shown in Fig.~\ref{fig1}, we perform Tersoff based molecular dynamics (MD) simulations via LAMMPS \cite{LAMMPS} on an initially prepared 96 atom zeolite silica structure \cite{DatabaseZeoliteStructures}. The annealing quenching cycle destroys the zeolite structure and generates silicate structures in the amorphous form, as illustrated in Fig.~\ref{fig1}A. A total of 1000 simulated annealing quenching cycles with different initial random seeds are used to generate statistically significant amorphous structures. The radial distribution function (RDF) of the generated amorphous silica is in agreement with the literature \cite{shock_response_silica}.

We introduce the logarithmic radial distribution function (log-RDF), defined as \( {g}(r) \to \Tilde{g}(v) \), to extend the{range} from \( r \in (0, \infty) \) to \( v \in (-\infty, \infty) \), as shown in Fig.~\ref{fig1}{B}:
\begin{equation}
    \Tilde{g}(v) \sim \sum_{j} \sum_{i \neq j} \delta[v - \ln(|\mathbf{x}_{i} - \mathbf{x}_{j}|)].
    \label{RDF}
\end{equation}

In the following sections, \( \Tilde{g}(v) \) will be referred to as the log-RDF. A Gaussian background, resulting from the finite simulation size, is subtracted from the log-RDF, as illustrated by the dotted curve in Fig.~\ref{fig1}{D}.

The logarithmic partial radial distribution function (log-pRDF) captures structural correlations between specific types of atoms. In the silica amorphous network, as shown in Fig.~\ref{fig1}{D}, there are three types of log-pRDFs: silicon-silicon (Si-Si), oxygen-oxygen (O-O), and silicon-oxygen (Si-O), represented by the green, blue, and red curves, respectively. Highlighted regions in Fig.~\ref{fig1}{D}(1-4) correspond to the first through fourth nearest neighbor distances.

While the amorphous network does not have long range order, the interstructural network in the bulk can be identified as a set of edge sharing closed loop structures with various structural anisotropy. Presented in Fig.~\ref{fig1}{A}, N member closed loop structures in the above simulated amorphous silica can be identified via N alternating oxygen (O) and silicon (Si) atoms, where the ending atom is identical to the initial atom. We refer to such structures as N member native ring structures (N-NRS). From this definition, N must be even to satisfy the bonding requirements: each nearest \( \rm Si O \) pair in each N-NRS has a covalent bond with length \( \rm d_{\rm sio} \approx 1.64 \, \text{\AA} \). Each N-NRS can be written as a collection of atomic coordinates \( [\mathbf{X}^N_{u}] \in \mathbb{R}^{3N} \),
\begin{equation}
[\mathbf{X}^N_{u}]= \begin{bmatrix}
&\mathbf{x_{1}}(u)\\
&\mathbf{x_{2}}(u)\\
&\dots\\
&\mathbf{x_{N}}(u)
\end{bmatrix}, \hspace{0.2cm } \braket{[\mathbf{X}^N_{u}]}_N= \frac{1}{N} \sum_{i=1}^N \mathbf{x_{i}}(u),
\end{equation}
where \( \braket{[\mathbf{X}^N_{u}]}_N \) is its center coordinate and \( u \) labels a specific N-NRS within the amorphous network, and all possible N-NRSs form the N-NRS set,
\begin{equation}
\mathcal{E}_N=\{[\mathbf{X}^N_{u}],[\mathbf{X}^N_{u'}],[\mathbf{X}^N_{u''}] \cdots \}.
\label{ensemble}
\end{equation}
For any two atoms at \( \mathbf{x_{i}}, \mathbf{x_{j}} \) in the bulk of an amorphous network, we argue that there exists at least one edge sharing N-NRS (or two distinct paths) that contains both atoms. The corresponding number statistics of primitive rings have been extensively studied in the literature via common shortest path analysis \cite{ LEROUX201070}. While the detailed number statistics depend on different criteria \cite{YUAN2002343}, in this work, to minimize the sampling bias, we perform an unrestricted counting method to quantify the number statistics \( \mathcal{D}_0(N) \) of edge sharing N-NRSs using the following sampling procedure: for each MD generated amorphous network, we randomly select an initial coordinate \( \mathbf{x}_i \) and stochastically walk N steps to a coordinate \( \mathbf{x}_j \). If \( |\mathbf{x}_i - \mathbf{x}_j| \approx d_{\rm sio} \), we declare the path as an N member native ring structure (N-NRS). Otherwise, we repeat the process until all amorphous networks are sampled. In agreement with similar results presented in the literature \cite{king1967ring,PhysRevB.44.4925}, the N-NRS number statistics \( \mathcal{D}_0(N) \) obtained from this unrestricted counting method reflect the fraction of edge sharing N-NRS occurrences in amorphous networks, which increases exponentially as N grows. As shown by the solid black curve in Fig.~\ref{fig1}C, ranging from \( N = 6 \to 18 \), the above analytic form is compared against the nonprimitive N-NRS number statistics, sampled numerically from MD generated amorphous silica networks via the above unrestricted counting method.

The amorphous network formation occurs as soon as the quenching step starts, as illustrated in Fig.~\ref{fig1}{A}: silicon and oxygen atoms start to form covalently bonded clusters. Let's consider a hypothetical scenario during an annealing quenching cycle, where the annealing step turns the zeolite silica into a homogeneous liquid with free floating silicon and oxygen atoms with a density ratio of 1:2. As the quenching step is measured by the inverse temperature \( \beta = 1/k_B T \) (where \( k_B \) is the Boltzmann constant), silicon and oxygen clusters spontaneously form and could expand into local structures. While silicon and oxygen are 4-fold and 2-fold coordinated, the thermal fluctuations lead to noninteger effective coordination numbers \( c_{\rm Si}(\beta) \) and \( c_{\rm O}(\beta) \), respectively, resulting in temperature dependent N-NRS number statistics \( \mathcal{D}[N, c_{\rm Si}, c_{\rm O}] \), as outlined in the Appendix. As the temperature approaches zero, the number statistics approach those discussed above, i.e., \( \mathcal{D} \to \mathcal{D}_0 \).

\section{\rom{3}. Native Ring structural anisotropy in the amorphous network}
The structural anisotropy and geometric correlation of edge sharing N-NRSs are characterized by the N-member log-RDF \( \Tilde{g}^N(v) \) and log-pRDFs, \( \Tilde{g}^N_{\rm sisi}(v), \Tilde{g}^N_{\rm sio}(v), \Tilde{g}^N_{\rm oo}(v) \), respectively.

\begin{figure*}[t!]
    \centering
    \includegraphics{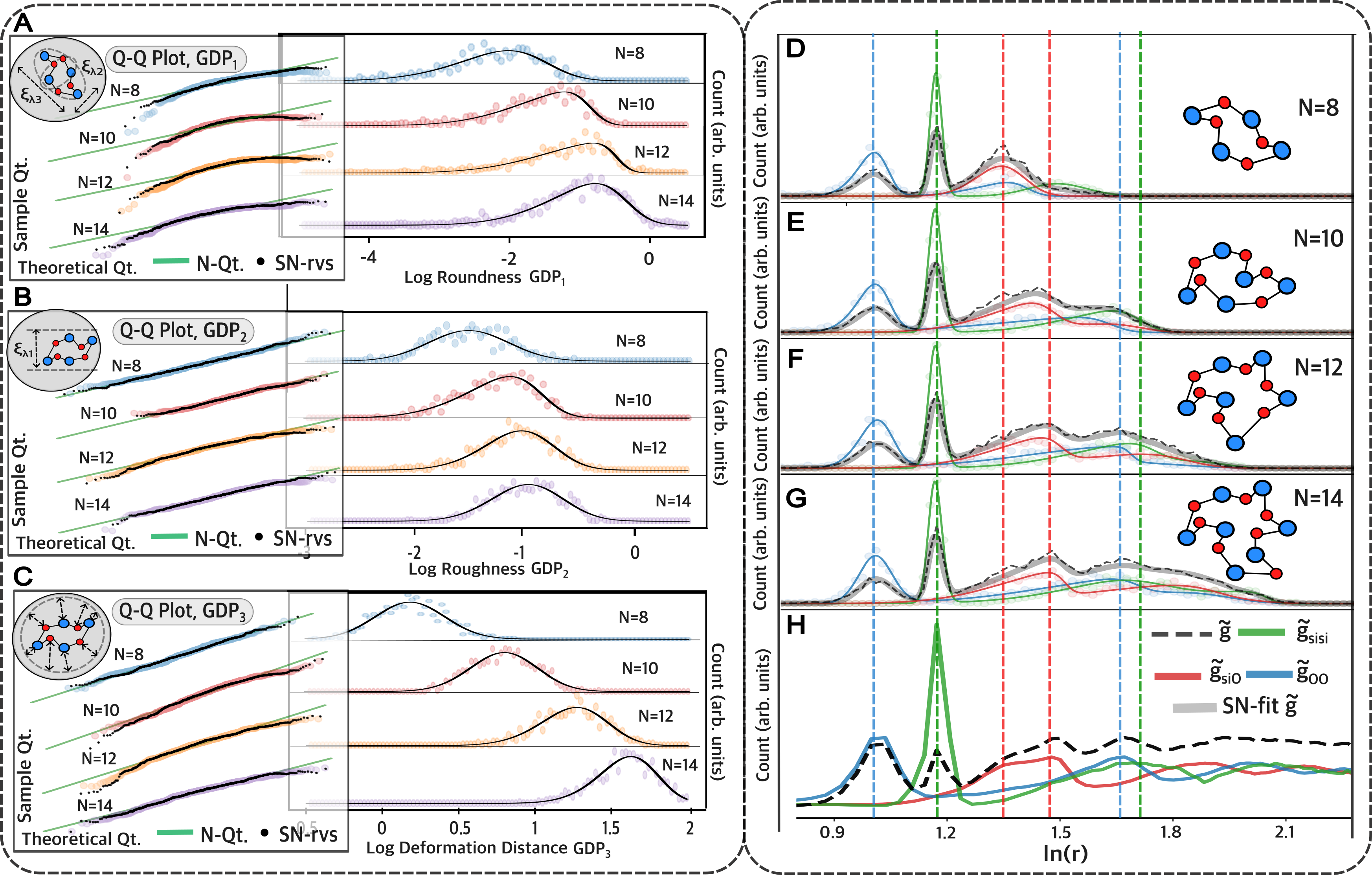}
    \caption{General anisotropy measures and global deformation parameters (GDP) are used to describe N-NRSs. We present three measures: roundness, roughness, and deformation distances, illustrated schematically in \textbf{A-C} inserts for N-NRSs with \( N = 8, 10, 12, 14 \). The corresponding GDPs, via the logarithm, are labeled as \( \rm GDP_1, GDP_2, GDP_3 \), respectively. Following the proof in Eq.~\ref{skew normal equation}, the three GDPs above follow skew normal (SN) distributions, regardless of detailed definitions. GDPs for N-NRSs with \( N = 8, 10, 12, 14 \) are shown with skew normal fittings. In Q-Q plots (left side of \textbf{A-C}), GDPs for all N-NRSs (identical color scheme) are compared against the skew normal random variable (black dots, SN-rvs), showing close numerical agreements. The green diagonal lines represent the referencing normal distribution. In \textbf{D-G}, the log-pRDFs of N-NRSs for each \( N = 8, 10, 12, 14 \) are presented and the log-pRDF of the silica amorphous network is shown in \textbf{H}. Across all \textbf{D-H},{we use reference vertical dotted lines to trace the log-RDF peaks, following the same color scheme as Silicon-Silicon (\( \Tilde{g}_{\mathrm{sisi}} \), green), Oxygen-Oxygen (\( \Tilde{g}_{\mathrm{oo}} \), blue), and Silicon-Oxygen (\( \Tilde{g}_{\mathrm{sio}} \), red). The total log-RDFs in each panel are shown as black dotted lines, along with the skew-normal (SN) fit \( \Tilde{g} \).
} These suggest that the anisotropy of the amorphous network is encoded by an ensemble of ring structures.
 }
    \label{fig2}
\end{figure*}

In addition, as an important type of anisotropy measure, we introduce the \( a \)-th nearest neighbor distance (graphically presented in Fig.~\ref{fig1} {D}), which measures the pair distances between two neighboring atoms. The above log-RDF and log-pRDF can be further decomposed into a sum of individual nearest neighbor distance contributions \( p_a(v) \),
\begin{equation}
    \Tilde{g}^N(v) \sim \sum_{a=1}^{N/2+1} {p}_a(v),  \hspace{0.3cm} {p}_a(v) \sim \sum_i \delta[v - \ln(r_{i,a})], 
    \label{n.n.}
\end{equation}
such that \( r_{i,a} = |\mathbf{x}_i - \mathbf{x}_{i+a}| \), where \( (i+a) \) has a modulus of N. This straightforward decomposition implies that the interconnectivity of an amorphous network is encoded in an ensemble of N-NRSs, and structural correlations and anisotropies can be inferred accordingly. As demonstrated in Fig.~\ref{fig2}(D-G), we present log-pRDFs for various N-NRSs, \( N = 8, 10, 12, 14 \), and identify peaks that are consistent (vertical lines with different colors) with the log-pRDF of the amorphous network in Fig.~\ref{fig2}H. The results suggest that the amorphous network anisotropy can be understood as contributions from ensembles of N-NRSs, sampled from the amorphous network.

In general, log-RDF and log-pRDF are not the only structural anisotropy measures for describing the variations in an amorphous network. Let's consider a generic anisotropy measure \( \Delta^{N}_u \) that reflects spatial correlations of N-NRSs, which are added in quadrature,
\begin{equation}
\left( \Delta^{N}_u \right)^2 = \sum_{i=1}^N \sum_{j=1}^N \cdots \sum_{k=1}^N f^2_N\left[\mathbf{x_{i}}(u), \mathbf{x_{j}}(u), \cdots, \mathbf{x_{k}}(u)\right],
\end{equation}
and the function \( f^2_N \) has a polynomial expansion that reflects general correlations between the N-NRS's spatial coordinates,
\begin{equation}
f_N\left[\mathbf{x_{i}}, \mathbf{x_{j}}, \cdots, \mathbf{x_{k}} \right] \sim \left[\mathbf{x_{i}} \cdot \mathbf{x_{j}}\right]^a \left[\mathbf{x_{i}} \cdot \mathbf{x_{k}}\right]^b \left[\mathbf{x_{j}} \cdot \mathbf{x_{k}}\right]^c \cdots.
\label{anisotropy measure def}
\end{equation}
We note that the anisotropy measure is positively defined, \( \Delta^N_u \geq 0 \). When \( \Delta^N_u = 0 \), the N-NRSs are considered isotropic by arbitrary definitions. We introduce three measures to quantify N-NRS structural anisotropy: roundness \( \Delta^{N,1}_u \), roughness \( \Delta^{N,2}_u \), and deformation distances{\( \Delta^{N,3}_u \)}, where the zero reference is taken to be an N-member circle with bond length \( \mathrm{d_{sio}} = 1.64 \, \text{\AA} \).

The roundness and roughness are frame independent anisotropy measures derived from the following pair correlation matrix \( \mathbf{M} \), obtained from the \( u \)-th N-NRS configuration \( [\mathbf{X}^N_{u}] \in \mathcal{E}_N \), capturing its structural correlation between three orthogonal directions, as demonstrated in Fig.~\ref{fig2}A,B graphical inserts, respectively. The matrix element has the form,
\begin{equation}
\mathbf{M}_{ab} = \sum_{i=1}^{N} \sum_{j \geq i}^{N} [\mathbf{x}_{i}]_a(u) \cdot [\mathbf{x}_{j}]_b(u),
\end{equation}
where \( a, b \) label the three orthogonal directions. The corresponding three eigenvalues \( \lambda_1(u) \leq \lambda_2(u) \leq \lambda_3(u) \) capture the spatial correlation of the N-NRS, and the corresponding eigenvectors define the semi principal axes of a triaxial ellipsoid, sharing the identical center of geometry with the N-NRS,
\begin{equation}
\frac{x^2}{\lambda_1(u)} + \frac{y^2}{\lambda_2(u)} + \frac{z^2}{\lambda_3(u)} = 1,
\end{equation}
where \( x, y, z \) are the directions set by the three eigenvectors of the correlation matrix \( \boldsymbol{\xi}_{\lambda_1}(u), \boldsymbol{\xi}_{\lambda_2}(u), \boldsymbol{\xi}_{\lambda_3}(u) \), as illustrated in Fig.~\ref{fig2}{A-C} inserts, respectively. We define the roundness and the roughness as,
\begin{equation}
\Delta^{N,1}_u = \left( \frac{\lambda_3(u)}{\lambda_2(u)} - 1 \right)^2, \hspace{0.3cm} \Delta^{N,2}_u = \frac{[\lambda_1(u)]^2}{\lambda_2(u) \lambda_3(u)},
\end{equation}
such that both roundness and roughness become zero in the reference case (circle).

The deformation distance \( \Delta^{N,3}_u \) for the \( u \)-th N-NRS, \( [\mathbf{X}^N_{u}] \in \mathcal{E}_N \), is defined as the quadrature mean distance with respect to the zero reference configuration (circle) with coordinates \( \mathbf{x}_i^0 \), as presented in Fig.~\ref{fig2}C graphical insert,
\begin{equation}
\Delta^{N,3}_u = \frac{1}{N} \sqrt{\sum_{i} \left|\mathbf{x}_i(u) - \mathbf{x}_i^0\right|^2}.
\end{equation}
We note that \( \Delta^{N,3}_u \) is a frame dependent measure: the relative choice of the N-NRS coordinates \( \mathbf{x}_i \) and the reference coordinates \( \mathbf{x}_i^0 \) can be arbitrarily chosen.

The structural properties of an amorphous network are captured by ensembles that contain all possible N-NRS within the network. In other words, N-NRS can be understood as a different physical basis, i.e., the corresponding statistics of structural anisotropy of the network can be decomposed into contributions from individual N-NRS ring statistics (in the ring structure basis).

As given in Eq.~\ref{ensemble}, each N-NRS within a given ensemble encodes structural information that is reflected in various anisotropy measures, such as those defined above. Similar to radial distribution functions (RDFs), taking the logarithm of anisotropy measures extends their domain from \( (0, \infty) \) to \( (-\infty, \infty) \). We refer to the logarithm of these anisotropy measures, \( \Delta^{N}_u \), as \textit{global deformation parameters} (GDPs). They are considered global because they incorporate information from all atoms within an N-NRS, and they are termed deformation parameters because structural anisotropy captures geometric deviation from a zero-reference configuration.

Specifically, we define the logarithm of the nearest-neighbor distance, roundness, roughness, and deformation distance as follows: \( \mathrm{GDP}_0 = v = \ln(r) \) shown in Fig.~\ref{fig4}E-H, \( \mathrm{GDP}_1 \) in Fig.~\ref{fig2}A, \( \mathrm{GDP}_2 \) in Fig.~\ref{fig2}B, and \( \mathrm{GDP}_3 \) in Fig.~\ref{fig2}C, respectively. For each log-anisotropy measure \( A \), the N-NRS ensemble \( \mathcal{E}_N \) is associated with the following anisotropy statistics \( \rho^N_A \),
\begin{equation}
\rho^N_A(\mathrm{GDP}) \sim \sum_{u}^{\rm all} \delta \left( \mathrm{GDP} - \ln \left[ \Delta_u^{N} \right] \right).
\label{GDP def}
\end{equation}

In Fig.~\ref{fig2}A-C, represented by colored data points, we show the GDP statistics associated with N-NRSs of size \( N = 8, 10, 12, 14 \), sampled from MD-generated amorphous silica networks. All three GDP distributions show close agreement with skew-normal distributions, indicated by black fitting curves. We also compare each GDP distribution (colored dots) against a standard normal distribution using quantile–quantile (Q–Q) plots, shown in the left insets of Fig.~\ref{fig2}A-C. In each plot, the standard normal distribution is represented by a green diagonal line, and sampled skew-normal random variables (black dots) are compared alongside the GDP data, showing close agreement. Note that the Q–Q plots are shifted horizontally for visual clarity.

While these four anisotropy measures capture distinct structural properties of N-NRSs, in the following sections we demonstrate, both mathematically and numerically, that under certain approximations, the corresponding statistics follow skew-normal distributions.

 \begin{figure*}[t!]
    \centering
    \includegraphics{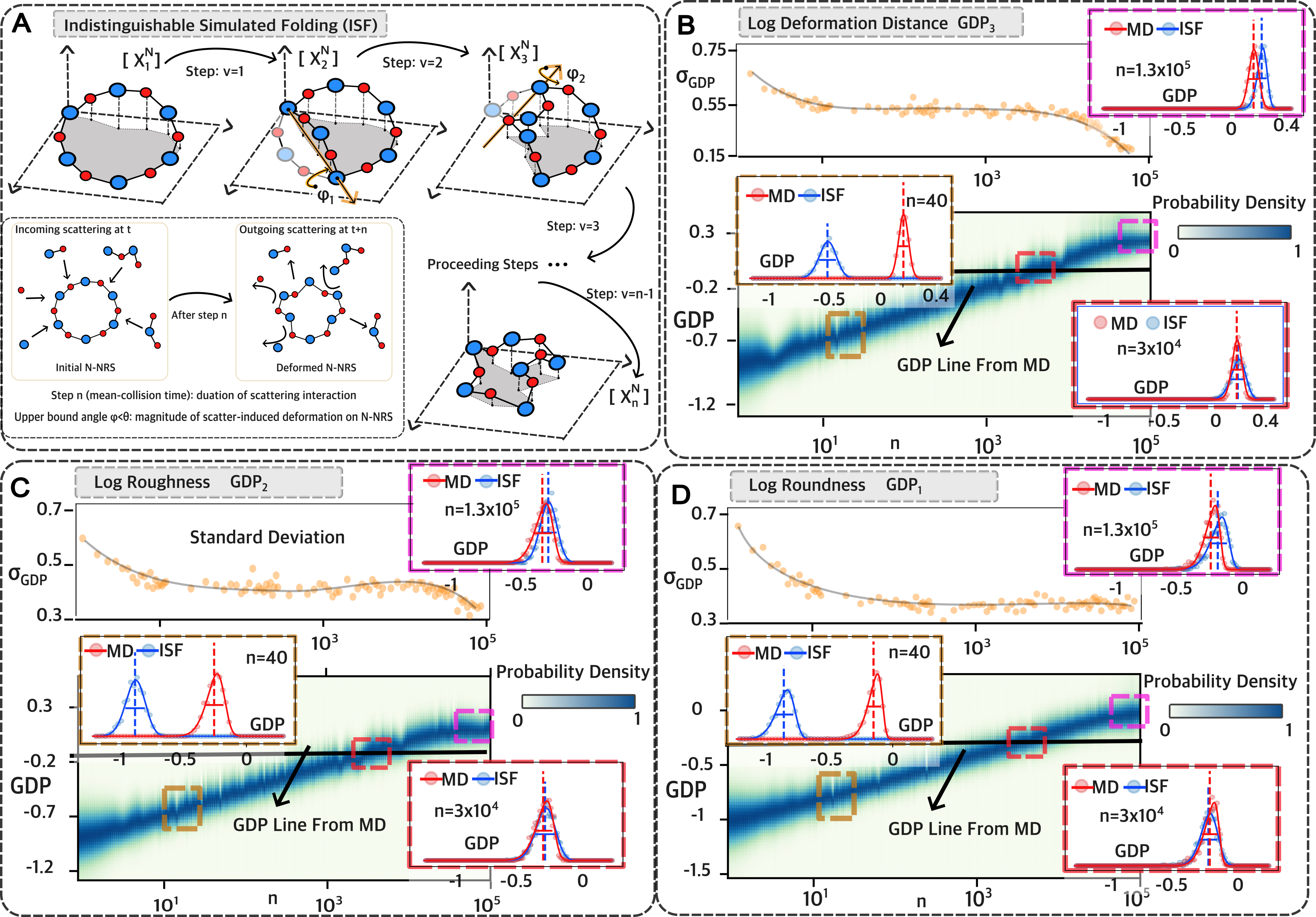}
    \caption{The concept of indistinguishable simulated folding (ISF). \textbf{A:} A graphical illustration of ISF. Starting from the initial ideal ring configuration, at each step, a rotation axis is randomly constructed by selecting a pair of atoms. Due to thermal fluctuations, a portion of atoms is rotated by a random angle \( \varphi_i \), bounded by the maximum angle \( \theta = 0.005\pi \). After N steps, the ISF process transforms the initial configuration into a final N-NRS. (\textbf{A}, inset) Schematic interpretation of the step parameter N and the upper bound angle \( \theta \) in the ISF framework. \textbf{B-D:} As defined in Eq.~\ref{GDP def}, the GDP statistics associated with three anisotropy measures, deformation distance, roughness, and roundness, are shown as heatmaps in panels \textbf{B}, \textbf{C}, and \textbf{D}, respectively. The standard deviations of the GDPs, \( \sigma_{\rm GDP} \), are plotted as orange data points. The black horizontal lines in each heatmap represent the expectation value of the GDP obtained from MD simulations. For each GDP heatmap, three vertical cross sections corresponding to folding steps \( n = 40 \), \( 3 \times 10^4 \), and \( 1.3 \times 10^5 \) are highlighted and compared with GDPs generated from MD simulations, as shown in the insets (outlined with matching colors). These vertical cross sections exhibit skew normal behavior in the GDPs (black curves), and this behavior remains consistent across the entire folding range \( n \in [1, 10^5] \), as described in Eq.~\ref{skew normal equation}. For each vertical slice, the N-NRS configurations are sampled from a set of 10-NRSs, with a sample size of 1000.
 }
    \label{fig3}
\end{figure*}

\section{\rom{4}. Concepts of Indistinguishable Simulated Folding (ISF) }
{It is important to determine how much of the structural anisotropy in an amorphous network arises from stochastic deformation accumulation during annealing and how much is induced by particle-specific many-body interactions. However, traditional direct simulation methods, such as MD simulations, are challenging because they simultaneously incorporate both many-body interactions and thermal fluctuations even in the simplest Lennard-Jones fluid model \cite{doi:10.1021/acs.jcim.9b00620}. So far, no method has been introduced to systematically distinguish these two contributions from each other.

The native ring structures within the network provide a convenient way to isolate and quantify the structural anisotropy arising solely from thermal fluctuations. }Let's consider a particular N-NRS within a local amorphous network during the quenching process, as illustrated in Fig.~\ref{fig3}A, with initial ideal ring configuration $[X_0^N]$, and at each discrete step, represented by $n'$, thermal fluctuations induce deformations on the N-NRS, leading to sequential changes in the N-NRS configuration from $n'=1$ to $n'=n-1$. Here, we assume that thermal fluctuations only activate relative rotational degrees of freedom (DOF) while preserving covalent bonds within the N-NRS. These covalent bonds also set the smallest length scale of the system. As demonstrated in Fig.~\ref{fig3}A, at each step $m$, the rotation axis $\mathbf{L}_{i,j}$, represented by an orange arrow, is defined by connecting two randomly selected atomic coordinates $\mathbf{x_i}(u)$ and $\mathbf{x_j}(u)$ within the N-NRS. The random angle of rotation $\varphi_{n'}$ reflects the magnitude of the thermal fluctuation, with a preset upper bound angle $\theta$. The SF sequence forms a Markov chain where the thermal fluctuation is assumed to have zero autocorrelation (memoryless) that the conditional probability of observing the next N-NRS configuration only depends on the present N-NRS configuration.

{While the initial ideal ring configuration $[X_0^N]$ might appear arbitrary and nonphysical, from a Monte Carlo perspective, a folding sequence can be viewed as a series of coordinate transformations that map an initial nonphysical ideal configuration onto a specific ring configuration considered {physical}. This concept is closely related to the Metropolis–Hastings algorithm used in sampling posterior (state) distributions: even if the initial proposal state is nonphysical, after a burn-in period, the process converges toward physical (stationary) states \cite{Metropolis1953} }

We define the bond-preserving folding sequence that transforms an initial N-NRS configuration into a final N-NRS configuration after N steps, $ [X_0^N] \to [X_n^N] $, as the \textit{simulated folding} (SF). This sequence simulates the physical process occurring during quenching: thermal fluctuations activate the rotational DOF of the N-NRS, generating deformations while the covalent structure remains approximately fixed. As the quenching process continues, the strength of thermal fluctuations decreases, leading to a reduction in the magnitude of structural deformations.{To eliminate nonphysical configurations, we implement a post-rejection sampling after completing the N-step Markov evolution, where the simulated configuration is rejected if pair distances are smaller than the smallest length scale $d_{\mathrm{sio}}$ (bond length). While the state space of the Markov chain is not subject to any artificially imposed constraint, we allow ``nonphysical'' intermediate configurations and post-select only those that satisfy physical criteria.}

We propose simulated folding as a minimal, constituent independent (particle independent) model that captures the stochastic structural deformation accumulated during the annealing-quenching cycle. The structural configurations of SF-generated N-NRSs statistically resemble the constituent independent portion of the N-NRS statistics produced by MD simulations. In the following sections, we consider a special case of SF, namely \textit{indistinguishable simulated folding} (ISF), where element types within the N-NRS are not distinguished, that is, silicon and oxygen are treated identically.{While constituent-dependent (many-body) interactions within a given length scale may be the dominant factor in generating structural deformations, which are not captured by the ISF model, we numerically demonstrate that, even without explicitly incorporating detailed many body interactions, the structural anisotropy statistics of ISF-generated N-NRSs capture the constituent-independent contributions of those observed in N-NRSs within the amorphous silica network.}

{Although it is intuitive to interpret step N in ISF as a continuous-time measure, it is important to clarify the distinction.} As illustrated in Fig.~\ref{fig3}(A, inset), the continuous-time interpretation of the step corresponds to the system's ``mean-collision time'' \( n(t) \) at a given conventional time \( t \); it reflects the duration of scattering events that generate deformations in the N-NRS. We refer to this ``mean-collision time'' as the ``step'' in the following sections, unless otherwise stated.

We note that the stochastic angle \( \varphi \), bounded by the upper limit \( \theta \), is related to the energy transfer rate from thermal fluctuations to inelastic structural deformations. For example, a larger upper bound angle corresponds to a greater unit of energy transfer.

Let us consider the rate of temperature loss, \( \frac{\partial T}{\partial t} \), resulting from the conversion of thermal fluctuations into inelastic structural deformations (neglecting other mechanisms that decrease the system temperature),
\begin{equation}
    \frac{\partial T}{\partial t} \sim -E[\theta, n] \, T, \hspace{0.2cm} T \sim \exp \left(-\int E[\theta, n] \, dt \right),
    \label{energy transfer}
\end{equation}
where \( \theta \) and N are implicitly time-dependent, and together define an energy scale \( E \). In other words, the rate of temperature change induced by the transfer of energy from thermal fluctuations to inelastic structural deformations (i.e., anisotropy) is governed by the rate \( E[\theta, n] \). To lowest order, assuming \( \theta \) and N are not explicitly correlated, we have,
\begin{equation}
    E[\theta,n] \sim -e_1 \frac{\partial \theta }{\partial t} + e_2 \frac{\partial n}{\partial t} + \cdots,
    \label{equation for real time}
\end{equation}
where \( e_1 \) and \( e_2 \) are positive scaling constants, and \( -\frac{\partial \theta}{\partial t}, \frac{\partial n}{\partial t} \geq 0 \). The first term captures the reduction in thermal energy transfer associated with decreasing deformation amplitude. The second term reflects changes in the mean-collision time, that is, the effective ``emptiness'' of the surroundings around a given atom.

{In a realistic system, the empirical cooling rate is a composite rate that is influenced not only by inelastic structural deformation and mean collision time, but also by other effects such as elastic deformation, heat convection through the environment, etc. Let us consider two idealized cases, the adiabatic and rapid regimes.} When thermal fluctuations are rapidly converted into structural deformations of the amorphous network, the first term dominates, \( E \sim -\frac{\partial \theta}{\partial t} \). We define this process as rapid quenching. On the other hand, when quenching is slow and the first term approaches zero, the dominant contribution becomes \( E \sim \frac{\partial n}{\partial t} \). We define this process as adiabatic quenching. When \( E = 0 \), the system is in thermal equilibrium, and the structural anisotropy statistics of the induced deformations remain time invariant, for example, those associated with the statistical profiles shown in Fig.~\ref{fig2}(A, B, C).

\section{\rom{5}. ISF sequence as a Markov Chain with a growing entropy production rate}

An ISF sequence is inherently irreversible and drives an initial N-NRS out of equilibrium. The increasing entropy production in ISF eventually brings the N-NRS to a steady state without reaching equilibrium with respect to the Markov step, similar to the out of equilibrium dynamics observed during quenching with respect to time. To explore this similarity, let us consider an N-step discrete time ISF sequence as a stochastic process evolving in the joint state space \( [\mathbf{X}^N_{n'}] \in \mathbb{R}^{3N} \). We begin with the ideal ring configuration \( [\mathbf{X}^N_0] \). The evolution of the N-NRS configuration can be expressed as a sequence of transformations \( \boldsymbol{T}(n', n'-1) \) such that,
\begin{equation}
    [\mathbf{X}^N_n] = \boldsymbol{T}(n, n-1)[\mathbf{X}^N_{n-1}] = \left( \prod_{n'=1}^{n} \boldsymbol{T}(n', n'-1) \right)[\mathbf{X}^N_0],
\end{equation}
where \( [\mathbf{X}^N_n] \) is the N-NRS configuration after N steps. Each transformation \( \boldsymbol{T}(n', n'-1) \) is determined by the random axes \( \mathbf{L}_{i,j} \) and rotation angles \( \varphi \leq \theta \) at step \( n' \), mimicking deformations generated by thermal fluctuations during quenching. In this work, without loss of generality, we use uniform distributions at each step in the ISF sequence,
\begin{equation}
    \mathbf{L}_{i,a} = \mathbf{x}_i - \mathbf{x}_a, \hspace{0.2cm} i, a \sim U(1, N), \hspace{0.2cm} \varphi \sim (0, \theta).
    \label{uniform}
\end{equation}

The ISF sequence, with respect to an initial ideal ring configuration \( [\mathbf{X}^N_0] \), \( [\mathbf{X}^N_1], [\mathbf{X}^N_2], \cdots, [\mathbf{X}^N_n] \), forms a Markov chain with marginal distributions \( \pi(\mathbf{X},1), \pi(\mathbf{X},2), \cdots, \pi(\mathbf{X},n) \), such that the conditional probability of observing the next N-NRS configuration depends only on the present N-NRS configuration,
\begin{equation}
    \mathrm{Pr}(\mathbf{X}, n \hspace{0.1cm}|\hspace{0.1cm} [\mathbf{X}^N_{n-1}], \dots, [\mathbf{X}^N_1]) = \mathrm{Pr}(\mathbf{X}, n \hspace{0.1cm}|\hspace{0.1cm} [\mathbf{X}^N_{n-1}], n-1).
\end{equation}
The evolution of the marginal probability follows accordingly:
\begin{equation}
\pi(\mathbf{X}, n) = \sum_{\mathbf{X}'} \pi(\mathbf{X}', n-1) \, \mathrm{Pr}\left( \mathbf{X}, n \hspace{0.1cm}|\hspace{0.1cm} \mathbf{X}', n-1 \right).
\end{equation}

We assume that the above Markov chain is irreducible and aperiodic, such that a non-equilibrium stationary (steady) state exists:
\begin{equation}
\sum_{\mathbf{X}'} \pi(\mathbf{X}', n-1) \, \mathrm{Pr}\left( \mathbf{X}, n \hspace{0.1cm}|\hspace{0.1cm} \mathbf{X}', n-1 \right) = \pi(\mathbf{X}, n),
\end{equation}
after \( n \to \infty \) steps. In contrast to an equilibrium state, where a stationary distribution satisfies detailed balance across all microstates, the non-equilibrium stationary distribution above arises from dynamic balance with a constant entropy production rate \cite{MARTYUSHEV20061}.

The Markov chain described above can be generalized to a continuous variable Markov chain. In the continuum limit, N becomes a continuous variable proportional to the mean collision time in an ISF sequence. The marginal probability then satisfies the following master equation (also referred to as a continuity equation) \cite{Landau_Binder_2014, pavliotis2014}:
\begin{equation}
\begin{aligned}
    R(\mathbf{X}, n) &= \frac{\partial_n \pi(\mathbf{X}, n)}{\pi(\mathbf{X}, n)} = \int_{V} dV_{\mathbf{X}'} \, \mathrm{\Tilde{Q}}(\mathbf{X}, n \hspace{0.1cm}|\hspace{0.1cm} \mathbf{X}', n) \\
    &\hspace{3cm} - \mathrm{Q}(\mathbf{X}', n \hspace{0.1cm}|\hspace{0.1cm} \mathbf{X}, n),
\end{aligned}
\label{continous pi}
\end{equation}

where \( R(\mathbf{X}, n) \) is the rate function for the local state \( \mathbf{X} \) at step N, \( V \) is the volume spanned by the N-NRS, and \( \Tilde{\mathrm{Q}} \) and \( \mathrm{Q} \) are time-dependent incoming and outgoing rate functions, respectively. We note that the incoming rate \( \Tilde{\mathrm{Q}} \) depends on prior steps, while \( \mathrm{Q} \) depends only on the current time. By summing the incoming \( \Tilde{\mathrm{Q}}(\mathbf{X}, n \hspace{0.1cm}|\hspace{0.1cm} \mathbf{X}', n) \) and outgoing \( \mathrm{Q}(\mathbf{X}', n \hspace{0.1cm}|\hspace{0.1cm} \mathbf{X}, n) \) rates, Eq.~\ref{continous pi} captures the cumulative rate of change for the local state \( \mathbf{X} \).

The first derivative of the rate function, \( \partial_n R(\mathbf{X}, n) \), characterizes the stability of transitions in the ISF sequence and is directly related to the change in the entropy production rate. If this derivative is negative, the system tends to relax toward thermal equilibrium by redistributing the Boltzmann weights across all microstates. In contrast, if the derivative is positive, the system is driven or dissipative, with energy continuously supplied or removed, pushing the system away from equilibrium toward a non-equilibrium steady state, where the weights of microstates remain dynamic.

In the ISF framework, the active folding of N-NRSs can be interpreted as a driven process in which the rate function satisfies the inequality \( \partial_n R(\mathbf{X}, n) \geq 0 \). This reflects the growth of the system’s entropy production rate until a non-equilibrium stationary distribution is reached. The inequality describes the direction of change in entropy production rate with respect to the mean collision time, rather than physical time, although the two are intrinsically connected.

We note that the magnitude of the rate \( R(\mathbf{X}, n) \) decreases over Markov steps according to the following power-law asymptotic {ansatz}:
\begin{equation}
   \lim_{n \to \infty} |R(\mathbf{X}, n)| \sim \frac{1}{f(n)} \cdots,
   \label{rate approx}
\end{equation}
where \( f(n) \) is a monotonically increasing function. In other words, when the entropy production rate stops growing, the rate itself approaches a constant value.

{It is important to distinguish between thermal equilibrium in physical quenching and the non-equilibrium steady state in an ISF sequence. Thermal equilibrium implies that both the Markov step N and the upper bound angle \( \theta \) are independent of time, whereas the non-equilibrium steady state in an ISF sequence is characterized by a constant entropy production rate with respect to the Markov step, \( \partial_n R(\mathbf{X}, n) \).}

 \section{\rom{6}. Non-Equilibrium Step evolution of general anisotropy measures in ISF sequences }
To quantify the structural anisotropy of the N-NRS associated with the ISF sequence, we analyze the changes in statistical measures of anisotropy as the Markov chain evolves. As previously discussed in Eqs.~\ref{energy transfer} and \ref{equation for real time}, while correlated, a linear evolution of the mean collision time (Markov step N) does not necessarily correspond to a linear progression in physical time.

Let us first examine the changes and accumulation of structural anisotropy statistics as the ``duration of scattering'' increases by considering a positively defined anisotropy measure, \( \Delta = G([\mathbf{X}]) \), associated with a mapping \( G: \mathbb{R}^{3N} \to \mathbb{R}^{>0} \). A general anisotropy measure may be scaled by an arbitrary power \( p \), such that \( (\Delta^N_n)^p \to \Delta^N_n \). The anisotropy measure is positively bounded, \( \Delta^N_n \geq 0^+ \), and it approaches \( 0^+ \) when there is no anisotropy. For example, an ideal ring structure (a perfect circle in two dimensions) has zero roughness.

As shown in the Appendix, the mean anisotropy measure \( \braket{G(\mathbf{X})}_n \) at step N, averaged over all N-NRS configurations, is given by,
\begin{equation}
\begin{aligned}
\Delta^N_n = \braket{G(\mathbf{X})}_n &= \int_{0}^{\infty} d\Delta \int_{V} dV_{\mathbf{X}} \delta(\Delta - G[\mathbf{X}]) \pi(\mathbf{X}, n) \\
&= \int_{V} dV_{\mathbf{X}} G(\mathbf{X}) \pi(\mathbf{X}, n).
\end{aligned}
\end{equation}

Here, the discrete step N is extended to a continuous representation such that \( \sum_n \to \int dn \). This corresponds to the intermediate steps within each folding operation, which are linearly proportional to the mean collision time. Under this continuation, the fractional change in the mean anisotropy measure \( \Delta^N_n \) from step N to \( n + dn \) is given by,
\begin{equation}
\begin{aligned}
\frac{\Delta^N_{n + \mathrm{d}n}}{\Delta^N_n}
&= \frac{\left[\Delta^N_n + \left( \partial_n \Delta^N_n \right) \mathrm{d}n + \cdots \right]}{\Delta^N_n} \\
&= 1 + \left( \frac{\braket{G(\mathbf{X}) R(\mathbf{X}, n)}_n}{\braket{G(\mathbf{X})}_n} \right) \mathrm{d}n + \mathcal{O}[(\mathrm{d}n)^2],
\end{aligned}
\label{change in anisotropy}
\end{equation}

where we relate the expectation value of the rate of change in the mean anisotropy measure to the rate function,
\begin{equation}
\partial_n \Delta^N_n = \braket{G(\mathbf{X}) R(\mathbf{X}, n)}_n = \int_{V} dV_{\mathbf{X}} R(\mathbf{X}, n) G(\mathbf{X}) \pi(\mathbf{X}, n),
\end{equation}

as demonstrated in the Appendix using Eq.~\ref{continous pi}.
The fractional change in the mean anisotropy measure generated via the infinitesimal step \( \mathrm{d}n \) in Eq.~\ref{change in anisotropy} can be written in an iterative form. Let us consider the stepwise evolution of the mean anisotropy measure resulting from the transition of the marginal distribution \( \pi(\mathbf{X}, n_0) \to \pi(\mathbf{X}, n) \),
\begin{equation}
\begin{aligned}
     \Delta^{N}_{n} &= \left(\frac{[\Delta^{N}_{n_0}] [\Delta^{N}_{n}]}{\Delta^{N}_{n - \mathrm{d}n}} \right) \left( \frac{\Delta^{N}_{n - \mathrm{d}n}}{\Delta^{N}_{n - 2\mathrm{d}n}} \right) \cdots \left( \frac{\Delta^{N}_{n_0 + \mathrm{d}n}}{\Delta^{N}_{n_0}} \right) \\
     &= \Delta^{N}_{n_0} \prod_{k=1}^{K} \left[ 1 + \left( \frac{\braket{G(\mathbf{X}) R(\mathbf{X}, n_k)}_{n_k}}{\braket{G(\mathbf{X})}_{n_k}} \right) \mathrm{d}n \right],
\end{aligned}
\end{equation}
where the number of infinitesimal foldings is given by the ratio \( K = (n - n_0) / \mathrm{d}n \). This product becomes a sum under the logarithm, and the mean global deformation parameter (GDP) is defined as the logarithm of the ratio between final and initial anisotropy measures,
\[
I(n_0, n) = \ln \left( \frac{\Delta^{N}_{n}}{\Delta^{N}_{n_0}} \right).
\]
To first order, using the approximation \( \ln(1 + \mathrm{d}n) \approx \mathrm{d}n \), we obtain the following integral form,
\begin{equation}
\begin{aligned}
    I(n_0, n) &= \int_{n_0}^{n} \mu_{n'} \, \mathrm{d}n' = \int_{n_0}^{n} \frac{\braket{G(\mathbf{X}) R(\mathbf{X}, n')}_{n'}}{\braket{G(\mathbf{X})}_{n'}} \, \mathrm{d}n',
\end{aligned}
\label{GDP mean}
\end{equation}
where we define the integrand as \( \mu_{n'} \). The resulting anisotropy measure \( \Delta^{N}_{n} \) can thus be expressed as an exponential function of the initial anisotropy \( \Delta^{N}_{n_0} \), similar to behavior in diffusion-driven systems \cite{Lawler_Limic_2010}, with deterministic contributions from the rate function at each intermediate step.

While the mean GDP evolves deterministically, individual GDPs fluctuate due to various random processes. Let us now consider a general case in which the master equation in Eq.~\ref{continous pi} evolves under a \textit{retarded} stochastic rate function \( R^{\mathcal{R}}(\mathbf{X}, n) \), perturbed by random fluctuations \( \gamma \) over a stochastic Markov step duration \( \epsilon_n \in \mathbb{R} \),
\begin{equation}
R^{\mathcal{R}}(\mathbf{X}, n) \to R(\mathbf{X}, n - \epsilon_n \, \mathrm{Sgn}[\epsilon_n]) + \mathrm{Sgn}[\epsilon_n] \int_0^{\epsilon_n} \gamma(\epsilon') \, d\epsilon',
\end{equation}
where we introduce the sign function \( \mathrm{Sgn} \) to preserve causality: contributions from advanced steps (\( \epsilon_n < 0 \)) are reversed. Intuitively, the two stochastic terms \( \epsilon_n \) and \( \gamma \) represent fluctuations in the mean collision time and in the rate function, respectively. The physical interpretation of the retarded stochastic rate function is as follows: the master equation evolves to a stochastic step \( n - \epsilon_n \) prior to the final step N with rate function \( R \), and the remaining portion of the evolution occurs stochastically over a duration \( \epsilon_n \), governed by \( \gamma(\epsilon) \), originating from step \( n - \epsilon_n \).
Eq.~\ref{GDP mean} becomes a stochastic equation, and the GDP becomes a \textit{random variable}, \( \mathbf{I}(n_0, n) \),
\begin{equation}
\begin{aligned}
\mathbf{I}(n_0, n) &= \int_{n_0}^{n} \left( \frac{\braket{G(\mathbf{X}) R^{\mathcal{R}}(\mathbf{X}, n')}_{n'}}{\braket{G(\mathbf{X})}_{n'}} \right) \mathrm{d}n'.
\end{aligned}
\label{GDP SE}
\end{equation}

As detailed in the Appendix, this random variable is skewed and can be written as the sum of three terms,
\begin{equation}
\mathbf{I}(n_0, n) = I(n_0, n) + \int_{n_0}^{n} \left( -\epsilon_{n'} \mathcal{A}_{n'} \, \mathrm{Sgn}[\epsilon_{n'}] + \Gamma_{n'} \right) \mathrm{d}n',
\label{int value}
\end{equation}
where \( \mu_n, \mathcal{A}_n, \Gamma_n \) are functions of \( \mathbf{I}(n_0, n) \). The first term captures the deterministic drift of the mean GDP. The second term represents a noise-induced drift that captures the retarded stochastic contribution associated with step uncertainty \( \epsilon_n \), where
\[
\mathcal{A}_n = \frac{\braket{G(\mathbf{X}) \partial_n R(\mathbf{X}, n)}_n}{\braket{G(\mathbf{X})}_n}.
\]
This term is directly related to the growth of the entropy production rate in a non-equilibrium process and satisfies \( \mathcal{A}_n \geq 0 \). The third term,
\[
\Gamma_n = \int_0^{\epsilon_n} \gamma_n(\epsilon') \, d\epsilon',
\]
is analogous to a diffusion term and reflects the cumulative stochastic contribution from thermal fluctuations. Without loss of generality, the sign function is omitted in the third term since it does not affect the range of \( \Gamma_n \).

We note that both stochastic terms are determined by the random axes \( \mathbf{L}_{i,j} \) and rotation angles \( \varphi \leq \theta \) at each step N, as described in Eq.~\ref{uniform}. The above discussion remains valid for any ISF-like Markov chain that exhibits a growing entropy production rate. While worthy of further exploration, Eq.~\ref{int value} can be extended into a stochastic differential equation for the GDP using a generalized Fokker–Planck equation \cite{Steeb1997, evans2012introduction, Chung1990}.

 \begin{figure*}[t!]
    \centering
    \includegraphics{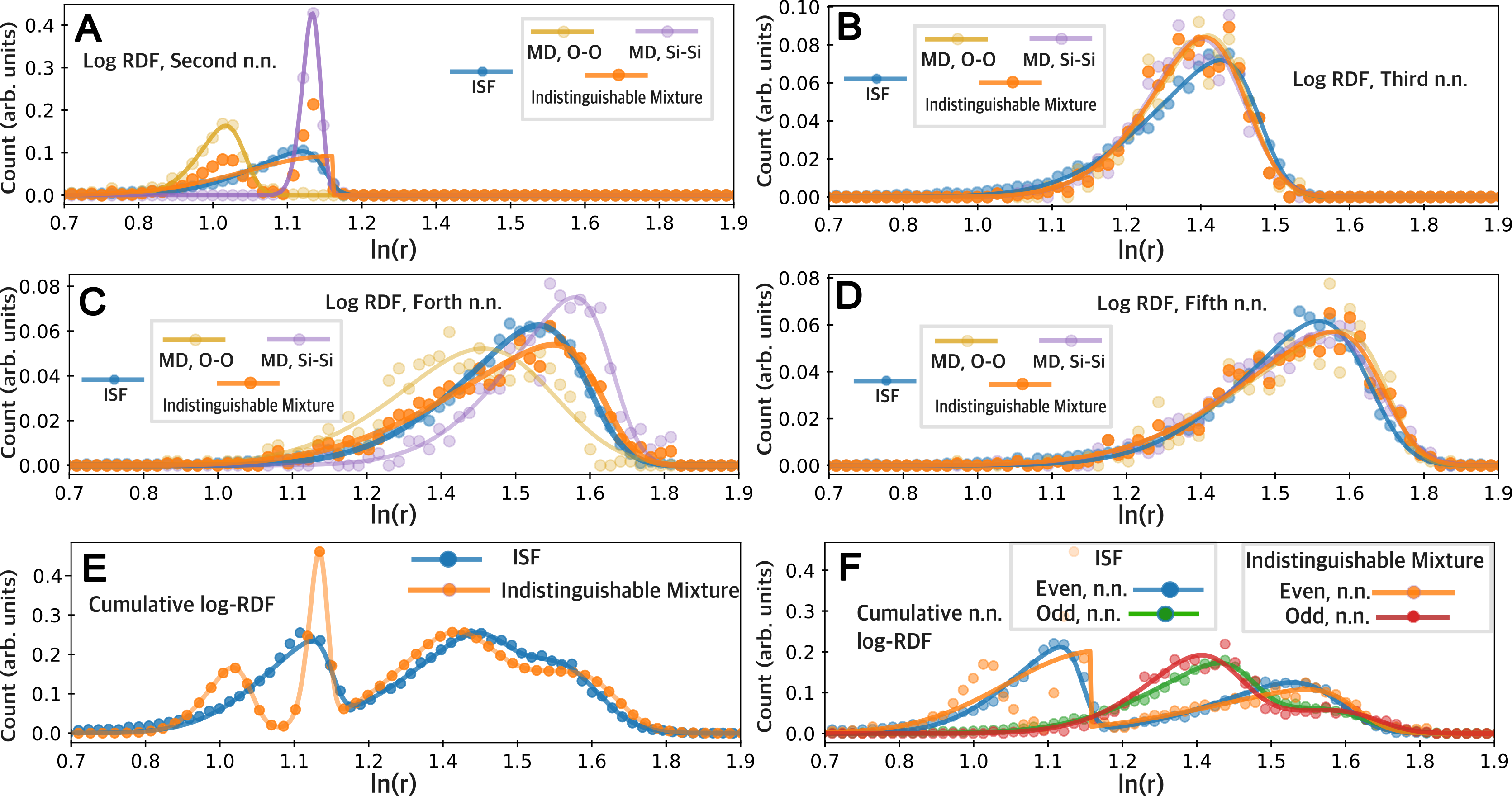}
    \caption{Nearest neighbor log-RDF contributions sampled from the MD N-NRS set (orange) and the ISF N-NRS set (blue), respectively. \textbf{A, C:} Even nearest neighbor log-RDF contributions. The indistinguishable mixture model (orange) represents an equal mixture of \( \Tilde{g}([\text{O-O}]) \) (yellow) and \( \Tilde{g}([\text{Si-Si}]) \) (purple), as defined in Eq.~\ref{mix}. 
\textbf{B, D:} Odd nearest neighbor log-RDF contributions. \textbf{E:} Cumulative log-RDFs for the ISF and MD N-NRS sets, reconstructed by summing all nearest neighbor contributions shown in \textbf{A-D}. \textbf{F:} Partial sums of odd nearest neighbor log-RDFs (red and green) and even nearest neighbor log-RDFs (orange and green). The red and orange curves are sampled from the MD N-NRS set, while the green and blue curves are sampled from the ISF N-NRS set, as indicated in the legend. The ISF log-RDF closely matches the constituent-independent portion of the indistinguishable mixture model. Nearest neighbor statistics are computed from 1000 N-NRSs with \( \theta = 0.005\pi \), \( n = 3 \times 10^4 \), sampled from the 10-NRS set. } 
    \label{fig4}
\end{figure*}

 \section{\rom{7}. Skew-normal Distribution: the GDP statistics driven by the Wiener process }
In addition to the expectation value \( I(n_0, n) \), the evolution of GDP, \( \mathbf{I}(n_0, n) \), as a random variable, includes two additional stochastic contributions: a stochastic drift term and a diffusion term, as shown in Eq.~\ref{int value}. The former remains non-negative at all steps, originating from the non-negative growth of the entropy production rate in an ISF process. The latter captures the stochastic contribution in the master equation, arising from uncertainty in the step duration, \( \epsilon_n \). To ensure causality in the stochastic term, we introduce the sign function \( \mathrm{Sgn}(\epsilon_n) \). 

In this section, we propose, with numerical examples, that when the above stochastic processes are modeled as Wiener processes, the GDP becomes a skew-normal random variable. 
{We numerically illustrate that, without incorporating any particle-specific interactions, the ISF-generated skew-normal statistics reflect the particle-independent portion of the MD-generated GDP statistics under specific mean collision times (step N) and upper bound angles \( \theta \).}

The stochastic drift term in Eq.~\ref{int value} remains non-negative, whereas the diffusion term can take both positive and negative values. The simplest processes that generate these stochastic terms are Wiener processes. If \( \mathcal{A}_{n'}, \Gamma_{n'} \) are slow-varying, then to the lowest order, we propose the following form:
\begin{equation}
\begin{aligned}
\mathbf{I}(n_0, n) &= I(n_0, n) - \mathcal{A}_n |W_{n - n_0}| + \Gamma_n \Tilde{W}_{n - n_0} + \cdots,
\end{aligned}
\end{equation}
where \( W_{n - n_0} \) and \( \Tilde{W}_{n - n_0} \) are independent Wiener processes with variance \( n - n_0 \). Under these assumptions, the GDP as a random variable follows a skew-normal distribution \cite{10.1093/biomet/83.4.715},
\begin{equation}
\mathbf{I}(n_0, n) = I(n_0, n) + \sigma_n \boldsymbol{\mathcal{S}}_n + \cdots,
\label{skew normal equation}
\end{equation}
where \( \boldsymbol{\mathcal{S}}_n \) is a standard skew-normal random variable with skew parameter determined by the ratio between \( \mathcal{A}_n \) and \( \Gamma_n \),
\begin{equation}
\boldsymbol{\mathcal{S}}_n \sim \mathcal{SN}\left(-\sqrt{\frac{\mathcal{A}_n}{\Gamma_n}}\right), \hspace{0.2cm} \sigma_n = \sqrt{(\mathcal{A}_n)^2 + (\Gamma_n)^2}.
\end{equation}
Eq.~\ref{skew normal equation} suggests a universal behavior of ISF-generated GDPs: regardless of the specific definition of GDP, to lowest order, the random variable obtained from an ISF sequence follows a skew-normal distribution.

We numerically demonstrate this skew-normal property via GDPs associated with roundness, roughness, and deformation distance of ISF-generated N-NRS sets with \( N = 8, 10, 12, 14 \), shown in Fig.~\ref{fig3} (B-D). Each vertical slice in the heatmap represents the GDP statistics from an N-step ISF sequence, where \( n \in [1, 10^6] \). At representative steps \( n = 40, 3 \times 10^4, 1.3 \times 10^5 \), GDP statistics generated via ISF (vertical slice) are compared to the GDP statistics sampled from MD simulations in Fig.~\ref{fig2}. To second order (first and second cumulants), close agreement (highlighted in red boxes) around step \( n = 3 \times 10^4 \) suggests correspondence in the anisotropy statistics between the ISF and MD-generated distributions.

In addition to the typical anisotropy measures discussed above, the \( a \)-th nearest neighbor distance also serves as a positively defined anisotropy measure. At step \( n = 3 \times 10^4 \) and upper bound angle \( \theta = 0.005\pi \), we further compare individual nearest neighbor log-RDFs of the N-NRS set (\( N = 10 \)), generated via ISF and MD simulations, in Fig.~\ref{fig4}. 
{Nearest neighbor distances in MD simulations are particle-specific, with distinct pair distributions for silicon-silicon (Si-Si, $d\sim  0.31 \mathrm{nm}$) and oxygen-oxygen (O-O, $d\sim  0.27\mathrm{nm}$) pairs, which display characteristic peaks \cite{10.1063/1.3632968,Mozzi:a07098}.}

 \begin{figure*}[t!]
    \centering
    \includegraphics{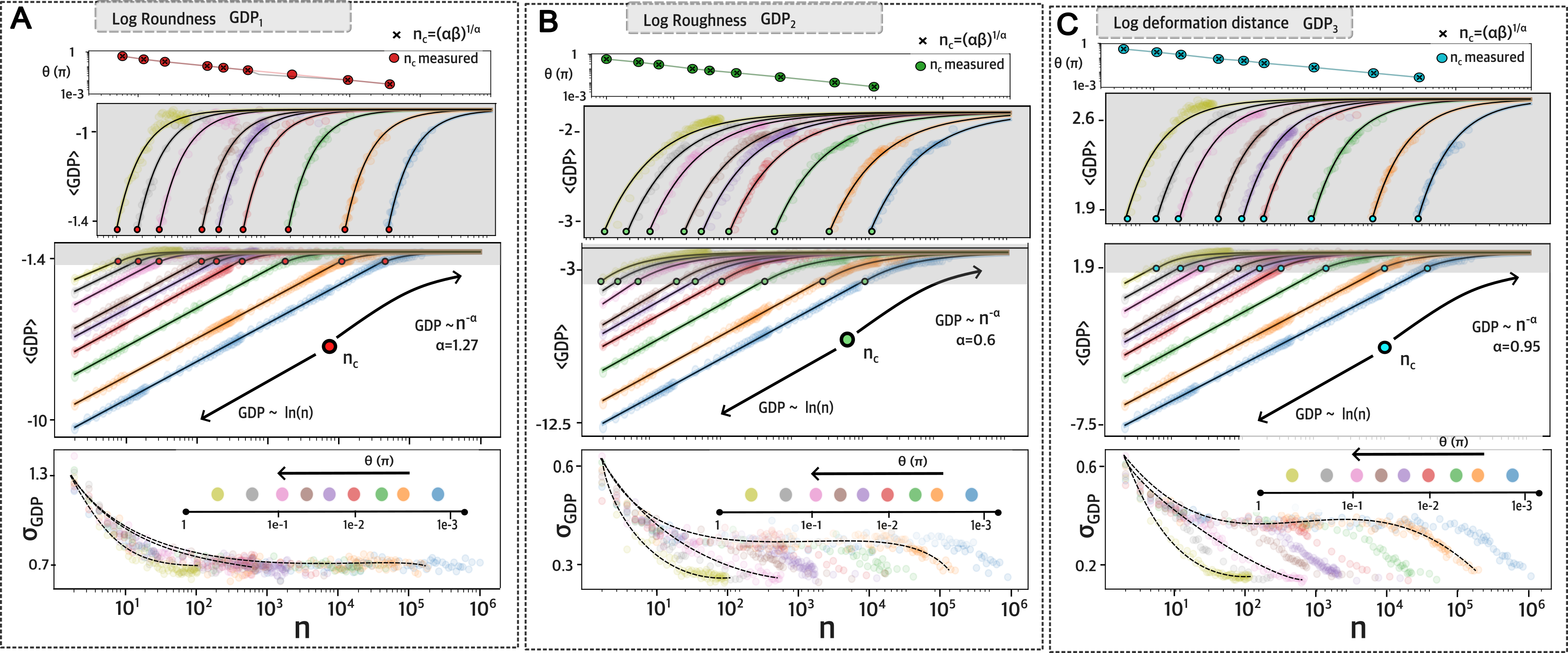}
    \caption{\textbf{A-C:} Expectation values of GDPs and the scaling laws given in Eq.~\ref{scaling laws}. When N is small, the expectation values of GDPs exhibit logarithmic divergence, such that as \( n \to 0 \), \( \text{GDP} \to -\infty \) (middle panels). In contrast, for large N, the expectation values follow a power-law behavior (top panels), asymptotically converging to a fixed value. As shown in Eq.~\ref{intersect}, the intersection between these two scaling regimes defines the critical step \( n_c \), which is both analytically calculated and numerically measured for all three GDPs at different upper bound angles \( \theta \). The corresponding data points for \( n_c \) are labeled in red, green, and cyan above the top panels. The bottom panels display the standard deviations of GDPs across various upper bound angles \( \theta \). Black dashed lines are included as visual guides. Expectation values and standard deviations for each GDP are computed using 1000 N-NRS configurations sampled from the 10-NRS set.} 
    \label{fig5}
\end{figure*}

{In amorphous silica, two types of atoms form three partial distances: $[\text{O-O}]$, $[\text{Si-Si}]$, and $[\text{Si-O}]$, each contributing distinct many-body features to the log-RDFs.

$[\text{Si-O}]$ corresponds to the \textit{odd} nearest neighbors. For \textit{even} nearest neighbors, the pair type depends on the starting atom—either $[\text{O-O}]$ or $[\text{Si-Si}]$. The constituent-independent contribution cannot reflect this atomic specificity.

To leading order, we approximate the {even} nearest neighbors using an indistinguishable mixture model (up to normalization), \begin{equation} 
\Tilde{g}(X) = \frac{1}{2} \Tilde{g}([\text{O-O}]) + \frac{1}{2} \Tilde{g}([\text{Si-Si}]), 
\label{mix} 
\end{equation}
where with $\Tilde{g}([\text{O-O}])$, $\Tilde{g}([\text{Si-Si}])$ are nearest neighbor log-RDFs. 

As shown in Fig.~\ref{fig4}, each structural anisotropy observation comes from randomly selecting $[\text{O-O}]$ (yellow) or $[\text{Si-Si}]$ (purple) with equal probability, i.e. indistinguishable. The resulting anisotropy statistics are given by the sum (orange).

For each nearest neighbor in an N-NRS set ($N=10$), we compare ISF- and MD-generated log-RDFs (Fig.\ref{fig4}A-D). Since the log-RDF is the sum of nearest-neighbor contributions (Eq.\ref{n.n.}), we reconstruct cumulative nearest neighbor and log-RDFs (Fig.\ref{fig4}E-F). As expected, particle-specific many-body features (Fig.\ref{fig4}A) are absent in the ISF model, while higher-order log-RDFs (Fig.~\ref{fig4}C-D) show close agreement. This reflects the decay of structural correlations with distance.

As shown in the Appendix, introducing local fluctuations and higher-order mixture models to the MD N-NRS set (Fig.\ref{fig7}) recovers full agreement with ISF predictions for $N=10$ (Fig.\ref{fig8}) and $N=12$ (Fig.~\ref{fig9}).

This agreement suggests that the ISF-captured constituent-independent contribution can be treated separately from detailed many-body interactions, and that quenching dynamics play a universal role in shaping amorphous structure.

We note that all nearest neighbor log-RDF statistics follow a skew-normal distribution, consistent with theory and other anisotropy statistics (Fig.~\ref{fig2}).

To further demonstrate the correspondence between ISF- and MD-generated N-NRS sets, we examine the variation of structural anisotropy statistics with respect to N. As shown in Fig.\ref{fig10}, we compute the shifts in the peak positions of the nearest neighbor log-RDFs, generated by both ISF and MD, across the range $N=8$ to $N=18$ (Fig.\ref{fig10}A-L). These shifts are then normalized and compared to the number of ISF steps (mean-collision time). The resulting peak variation per N, shown in Fig.~\ref{fig11}, exhibits close agreement as the neighbor distance increases. This supports the interpretation that long-range pairs become uncorrelated and constituent-independent, consistent with the ISF framework.

We exclude the second nearest neighbor log-RDF, since O-O and Si-Si many-body peaks remain the dominant features in amorphous networks (Fig.~\ref{fig12}).

}

 \begin{figure*}[t!]
    \centering
    \includegraphics{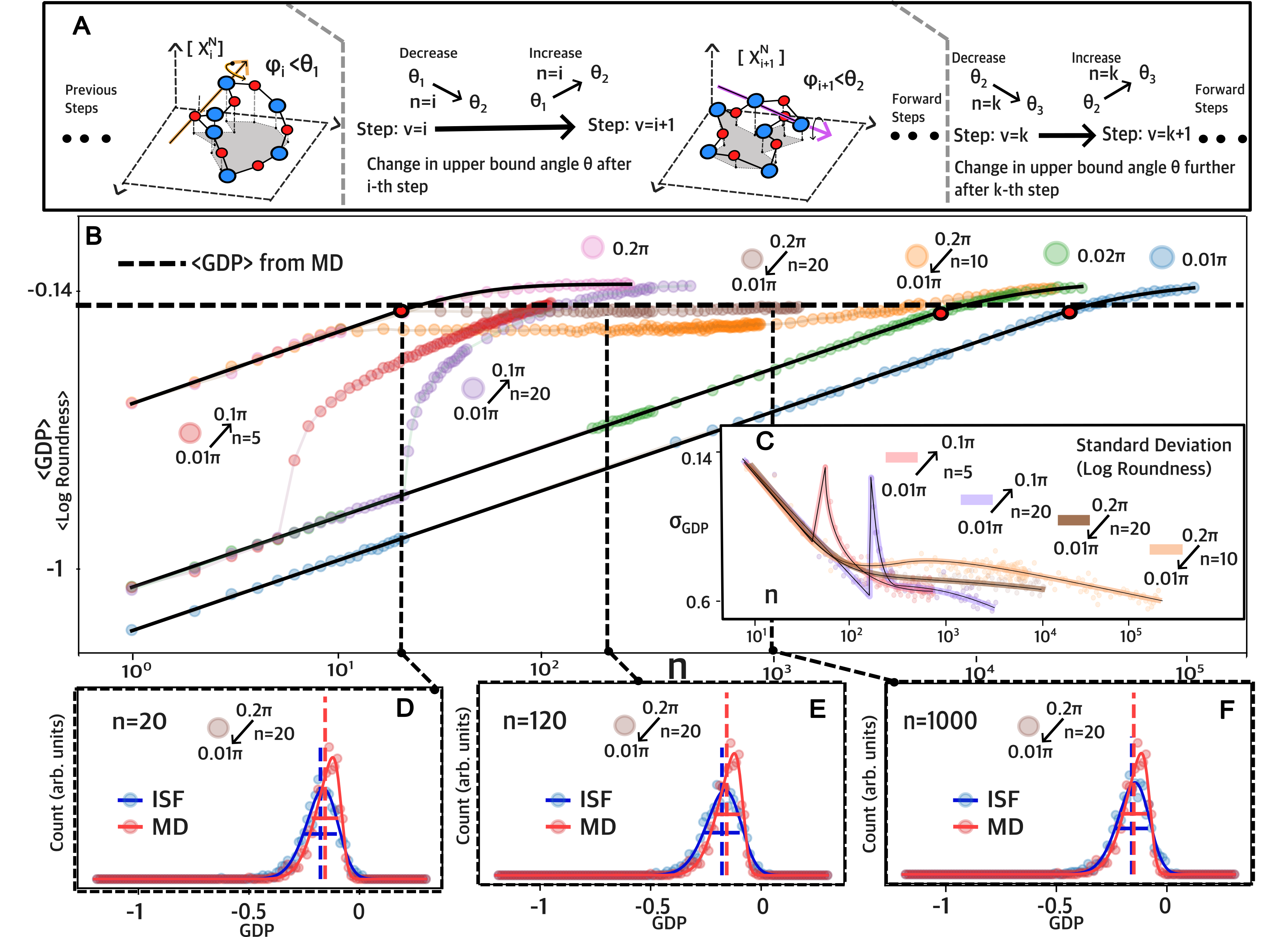}
    \caption{\textbf{A:} Illustration of the concept of rapid quenching in the ISF framework. \textbf{B:} Following the concept introduced in \textbf{A}, the ISF performs a rapid change in the upper bound angle from \( \theta_1 \) to \( \theta_2 \), indicated by the arrow and respective \( \theta \) values. While an annealing sequence (red and purple dots) leads to a significant increase in the expectation value of GDP (log roundness), a quenching sequence (orange and brown dots) stabilizes the GDP over an extended range of steps. This stabilization exceeds the rate of change observed in any adiabatic \( \theta \) sequence (blue and green curves). \textbf{C:} The standard deviation \( \sigma_{\rm GDP} \) of the corresponding quenching sequences. An incline sequence produces sharp spikes in \( \sigma_{\rm GDP} \), which quickly decay. In contrast, for a quenching sequence, \( \sigma_{\rm GDP} \) evolves more gradually compared to the adiabatic quenching scenario. \textbf{D-F:} The stabilization behavior of rapid quenching is further visualized in the GDP statistics at different ISF steps, \( n = 20, 120, 1000 \), respectively.
} 
    \label{fig6}
\end{figure*}

 \section{\rom{8}. Asymptotic Scaling}
In this section, we demonstrate the asymptotic behavior of the expectation value of a GDP. When the ISF step N exceeds a critical value \( n_c \), the expectation value of the GDP approaches a terminal value, obtained from the stationary distribution \( \pi_\infty \), such that \( \Delta_{n_0} = \Delta_n \). Using Eq.~\ref{GDP mean} and the ansatz in Eq.~\ref{rate approx}, we obtain,
\begin{equation}
\begin{aligned}
\lim_{n_0, n \gg n_c} I(n_0, n) \sim \int_{n_0}^{n} \frac{1}{f(n')} \, \mathrm{d}n' \to 0,
\end{aligned}
\end{equation}
for all \( n_0, n \gg n_c \). To the lowest order, we have the scaling,
\begin{equation}
f(n) \sim \frac{1}{\beta} n^{\alpha + 1}, \hspace{0.5cm} I(n_0, n) \sim -\beta n^{-\alpha},
\label{scale high}
\end{equation}
where the power \( \alpha \) is non-negative (\( \alpha > 0 \)) and \( \beta \) is a coefficient.

On the other hand, when \( n, n_0 \ll n_c \), the anisotropy measures approach zero, and we obtain,
\begin{equation}
\begin{aligned}
\lim_{n_0, n \ll n_c} I(n_0, n) \sim \int_{n_0}^{n} \frac{1}{f(n')} \, \mathrm{d}n' \to -\infty,
\end{aligned}
\end{equation}
with the expectation value of the GDP diverging logarithmically,
\begin{equation}
f(n) \sim n, \hspace{0.5cm} I(n_0, n) \sim \ln(n) + g,
\label{scaling laws}
\end{equation}
where \( g \) is an arbitrary constant.

In Fig.~\ref{fig5}A-C (middle panels), we compare the expectation values of three GDPs for various upper bound angles \( \theta \in [10^{-3}, 1] \), in units of \( \pi \), across the range \( n \in [1, 10^6] \). The shaded regions in the top panels correspond to those in the middle panels. The bottom panels show the standard deviations associated with each GDP. For large \( n \gg n_c \), the expectation values follow a power-law decay as described in Eq.~\ref{scale high}, illustrated in Fig.~\ref{fig5}A-C (top panels). In contrast, for small \( n \ll n_c \), the expectation values exhibit logarithmic divergence, as described in Eq.~\ref{scaling laws}, and shown in the middle panels.

The critical step \( n_c \) can be approximated by equating the two asymptotic expressions and their first derivatives with respect to N,
\begin{equation}
-\beta n_c^{-\alpha} = \ln(n_c) + g, \hspace{0.5cm} \frac{1}{n_c} = \alpha \beta n_c^{-\alpha - 1},
\label{intersect}
\end{equation}
which yields the analytical estimate,
\[
n_c = (\alpha \beta)^{1/\alpha}.
\]
As shown in Fig.~\ref{fig5}A-C (above top panels), this analytical estimate is compared with the numerically determined values of \( n_c \), revealing close agreement.

 \section{\rom{9}. adiabatic quenching and Rapid quenching  }
A close agreement between ISF-generated and{constituent-independent contributions} of MD-generated N-NRSs suggests that non-equilibrium dynamics plays a critical role in shaping amorphous network structures. As discussed in Eq.~\ref{energy transfer}, both the step N and upper bound angle \( \theta \) are directly correlated with the mean collision time and the local energy transfer rate. The rate of thermal fluctuations continuously induces deformations in the N-NRSs within the amorphous network. In this work, we consider two edge cases: adiabatic and rapid quenching.

During adiabatic quenching, the system maintains a near-zero rate of change in the upper bound angle \( \theta \) and accumulates thermally induced deformations by increasing the mean collision time N until the quenching process is completed. In contrast, in the rapid quenching regime, the rate of change in the upper bound angle \( \theta \) dominates, resulting in the system being locked into metastable configurations.

While in Figs.~\ref{fig3}, \ref{fig4}, and \ref{fig5}, numerical demonstrations are limited to the adiabatic quenching regime, Fig.~\ref{fig6} illustrates the rapid quenching regime, where the upper bound angle \( \theta \) becomes time-dependent. As shown in Fig.~\ref{fig6}A, during the early stages of quenching, thermal fluctuations are large in magnitude, leading to significant energy transfer and large N-NRS structural deformations, characterized by a large upper bound angle \( \theta_1 \). Since the system has a low coordination number, as defined in Eq.~\ref{corrdination number}, atoms are surrounded by more "empty space", corresponding to a small step N, i.e., a short mean collision time. As quenching continues, the coordination number increases, reducing the average empty surroundings. This increases the probability of collisions and leads to an increase in step N. Simultaneously, the decrease in temperature and local strain limits the extent of structural deformations from thermal fluctuations, resulting in a smaller upper bound angle \( \theta_2 \).

We model this behavior using a delta quenching sequence: starting with an upper bound angle \( \theta_1 \), after N steps, the angle is reduced to \( \theta_2 \) (\( \theta_1 \to \theta_2 \)), while the step N continues to evolve. This process can be repeated as a sequence, \( \cdots \to \theta_k \to \cdots \). Physically, under decreasing temperature, the N-NRS progressively undergoes less deformation until the quenching completes.

This concept is demonstrated numerically in Fig.~\ref{fig6}B-C for various quenching sequences. Notably, the brown data points in Fig.~\ref{fig6}B, D-F show that, unlike the curve for a single angle \( \theta = 0.01\pi \), the quenching sequence \( 0.2\pi \to 0.01\pi \) after 20 steps retains its GDP statistics over a wide range of subsequent N. This result reveals that in the rapid quenching regime, a large reduction in the energy transfer rate locks the N-NRS structures into an "intermediate state" that becomes insensitive to the duration of scattering events (step N). In other words, rapid quenching can cause the system to become trapped in a metastable state \cite{cohen1964metastability}. The rate of change in GDP statistics is proportional to the rate of temperature loss \( E[\theta, n] \), which reflects the net energy transfer from thermal fluctuations to inelastic structural deformations, as described in Eq.~\ref{energy transfer}:
\begin{equation}
    \frac{\partial I_{n}^N}{\partial t} = \frac{\partial \theta}{\partial t} \frac{\partial I_{n}^N}{\partial \theta} \sim -E[\theta, n] \frac{\partial I_{n}^N}{\partial \theta}.
\end{equation}

{We also recall that although the initial ideal ring configuration may seem nonphysical, the resulting N-NRS after a sequence of ISF steps can represent a physically meaningful structure.}

 \section{\rom{10}. Discussion and Conclusion}
Amorphous structures serve as critical hosts for next-generation quantum devices. In this work, we propose that the structural anisotropy of an amorphous network can be recovered through an ensemble of edge-sharing N-member native ring structures (N-NRSs) within the same network. Using classical molecular dynamics simulations, we sample N-NRSs from a set of MD-generated amorphous networks and quantify the corresponding statistics for each N. We demonstrate the correspondence of structural anisotropy between various N-NRS ensembles and the full amorphous network using log partial radial distribution functions (log-pRDFs). The structural anisotropy of N-NRSs is further quantified via the logarithm of any general positively defined anisotropy measure, as defined in Eq.~\ref{anisotropy measure def}, including four distinct measures introduced for numerical demonstration: roundness, roughness, deformation distance, and the \( a \)-th nearest-neighbor pair correlation function.

In the second part of this work, we introduce {indistinguishable simulated folding} (ISF), a non-equilibrium process with a growing entropy production rate that captures the constituent-independent contribution to structural anisotropy statistics within amorphous networks. ISF is formulated as a Markov chain that generates a stochastic deformation sequence, transforming an initial reference N-member ring configuration into an ISF-generated N-NRS. The step N, interpreted as the mean collision time, represents the duration of each scattering process that induces inelastic deformations in the initial configuration. At each step, the stochastic deformation is bounded by an upper limit on energy transfer from thermal fluctuations. Depending on the quenching conditions, and without introducing any prescribed interactions, we demonstrate that the structural anisotropy statistics of ISF-generated N-NRSs match the{constituent-independent contribution} of MD-generated N-NRS statistics, which are directly linked to the quenching rate.

In the adiabatic quenching regime, we show that the mean of a general global deformation parameter (GDP), defined as the logarithm of a positively defined anisotropy measure, is directly related to entropy production. The uncertainty in the GDP step evolution is linked to the non-negative entropy production rate, which introduces skewness in the GDP distribution. We prove that under Wiener process assumptions, any GDP random variable follows a universal skew-normal distribution.{While we propose the ISF framework to explain the constituent-independent contribution to structural anisotropy statistics of amorphous networks, it can be directly extended to deformable ring polymers, composite structures, and lattice models \cite{doi:10.1021/acs.macromol.5b00603,PhysRevMaterials.8.045601,PhysRevLett.57.3023}.}

In the final part of this work, we investigate GDP behavior in the rapid quenching regime and discover that rapid quenching enables metastable N-NRS configurations with persistent structural anisotropy, which remain stable under low-magnitude thermal fluctuations across extended steps. These findings highlight the robustness of the ISF framework in capturing key anisotropy features across diverse quenching regimes, offering a versatile tool for understanding ring deformation dynamics in amorphous systems.

\section{Acknowledgement}
We acknowledge support from NSF Award No. 2427169, 2137740 and Q-AMASE-i, through Grant No. DMR-1906325, and from NWO Quantum Software Consortium (Grant No. 024.003.037). Furthermore, the authors thank Yang Liu for her insightful discussions and suggestions. 

\section{Data Availability Statement}
The source code, sample data, and demo that support the findings of this study are openly available \cite{wang2025github}.

\medskip

\clearpage
\widetext

\section{Appendix}
\textbf{Noninteger coordination number and temperature dependence} has the following functional dependence:
\begin{equation}
\begin{aligned}
    \frac{\mathcal{D}[N,c_{\rm Si}, c_{\rm O}]}{\mathcal{D}_0(N)} &\sim \int_{0}^{\beta} \exp\left(-\beta' F[N,c_{\rm Si},c_{\rm O},\beta']  \right) d \beta',
\end{aligned}
\label{corrdination number}
\end{equation}
where \( F[N,c_{\rm Si},c_{\rm O},\beta] \geq 0 \) sets an energy scale that corresponds to each N-NRS formation type.

\textbf{General solution of the master equation} As mentioned in the main text, the general solution of the ISF master equation is given by the marginal probability \( \pi(\mathbf{X},n) \),
\begin{equation}
\begin{aligned}
\pi(\mathbf{X},n) = \pi(\mathbf{X},n_0)\exp\left( \int_{n_0}^n dn' R(\mathbf{X},n') \right) 
= \pi(\mathbf{X},n_0)\exp\left( \int_{n_0}^n dn' \int dV_\mathbf{X'} \left[ \mathrm{\Tilde{Q}}\left(\mathbf{X},n'\hspace{0.1cm}|\hspace{0.1cm} \mathbf{X}',n'\right)
- \mathrm{Q}\left(\mathbf{X}',n'\hspace{0.1cm}|\hspace{0.1cm} \mathbf{X},n'\right) \right] \right),
\end{aligned}
\end{equation}
with normalization,
\begin{equation}
    \int \pi(\mathbf{X},n)dV_{\mathbf{X}} = 1.
\end{equation}

\textbf{Variation in structural anisotropy measure} As discussed in the main text, the variation of a structural anisotropy measure can be expressed via the rate function \( R(\mathbf{X},n) \),
\begin{equation}
\begin{aligned}
\left[\partial_n\Delta^{N}_{n}\right] 
= \int dV_{\mathbf{X}} G(\mathbf{X}) \partial_n \pi(\mathbf{X},n) 
= \int dV_{\mathbf{X}} G(\mathbf{X}) R(\mathbf{X},n) \pi(\mathbf{X},n).
\end{aligned}
\end{equation}

\textbf{Stochastic equation for GDP} As discussed in the main text, the GDP becomes a random variable with the following stochastic equation:
\begin{equation}
\begin{aligned}
\mathrm{ln}\left(\frac{\Delta_n^N}{\Delta_{n_0}^N}\right)
&= \int_{n_0}^{n} \left( \frac{\braket{G(\mathbf{X}) R^{\mathcal{R}}(\mathbf{X},n')}_{n'}}{\braket{G(\mathbf{X})}_{n'}} \right) \mathrm{d}n' \\
&= \int_{n_0}^{n} \left( \frac{\braket{G(\mathbf{X}) \left[R(\mathbf{X},n' - \epsilon_{n'} \mathrm{Sgn}[\epsilon_{n'}]) + \mathrm{Sgn}[\epsilon_{n'}] \int_{0}^{\epsilon_{n'}} \gamma_{n'}(\epsilon') d\epsilon' \right]}_{n'}}{\braket{G(\mathbf{X})}_{n'}} \right) \mathrm{d}n' \\
&= \int_{n_0}^{n} \left( \frac{\braket{G(\mathbf{X}) R(\mathbf{X},n')}_{n'} 
- \braket{G(\mathbf{X}) \partial_{n'} R(\mathbf{X},n') \epsilon_{n'} \mathrm{Sgn}[\epsilon_{n'}]}_{n'} 
+ \braket{G(\mathbf{X}) \left( \mathrm{Sgn}[\epsilon_{n'}] \int_{0}^{\epsilon_{n'}} \gamma_{n'}(\epsilon') d\epsilon' \right)}_{n'}}{\braket{G(\mathbf{X})}_{n'}} \right) \mathrm{d}n' \\
&= \int_{n_0}^{n} \left( \mu_{n'} + \left( -\epsilon_{n'} \mathcal{A}_{n'} + \Gamma_{n'} \right) \mathrm{Sgn}[\epsilon_{n'}] \right) \mathrm{d}n',
\end{aligned}
\end{equation}
such that an increasing entropy production rate \( \partial_{n'} R(\mathbf{X},n') \geq 0 \) enforces \( \mathcal{A}_{n'} \geq 0 \), as discussed in the main text. The corresponding differential form is given by:
\begin{equation}
\partial \mathbf{I}(n_0,n) = \mu_n \mathrm{d}n + \left( -\mathcal{A}_n \epsilon_n \mathrm{d}n + \Gamma_n \mathrm{d}n \right) \mathrm{Sgn}[\epsilon_n].
\end{equation}

 \begin{figure*}[t!]
    \centering
    \includegraphics{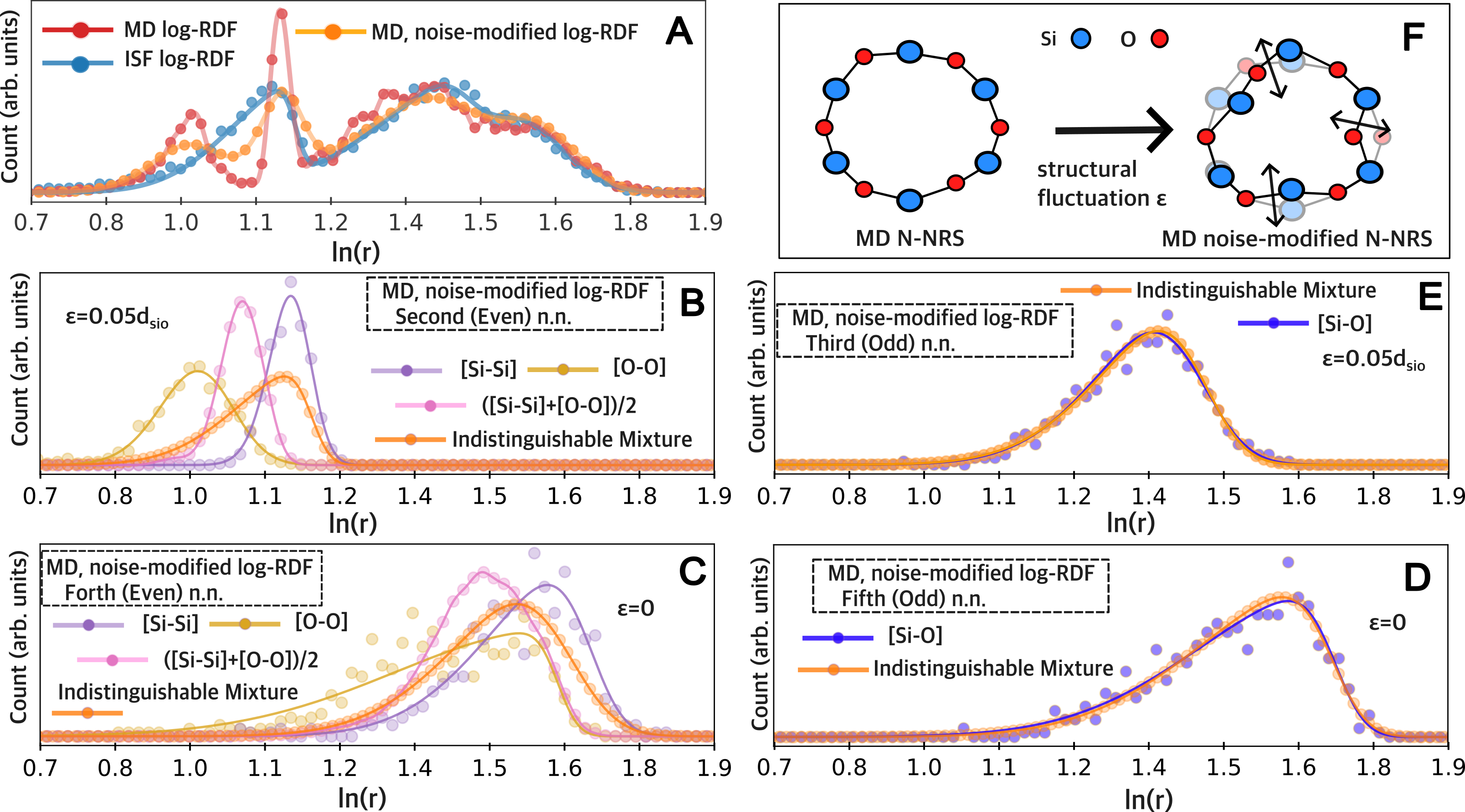}
    \caption{{Concept of Noise-Modified N-NRS Set and Indistinguishable Mixture Model.} 
\textbf{A:} Comparison of log-RDFs obtained from the MD-generated (red), noise-modified MD (orange), and ISF-generated N-NRS sets. 
\textbf{B-C:} Log-RDFs of the even nearest neighbors (n.n.) with structural fluctuations $\epsilon = 0.05\,\mathrm{d_{iso}}$ and $\epsilon = 0$. As discussed in the main text, the indistinguishable mixture model (orange) is a convex combination of three nearest neighbor log-RDFs: \( \Tilde{g}([\text{O-O}]) \) (yellow), \( \Tilde{g}([\text{Si-Si}]) \) (purple), and \( \Tilde{g}([C]) \) (pink). 
\textbf{D-E:} Log-RDFs of the odd nearest neighbors (n.n.) with structural fluctuations $\epsilon = 0.05\,\mathrm{d_{iso}}$ and $\epsilon = 0$, which uniquely define \( \Tilde{g}([\text{Si-O}]) \). 
\textbf{F:} Concept of structural fluctuation, where each original MD-generated N-NRS is modified by slightly perturbing atomic positions around their original coordinates.
} 
    \label{fig7}
\end{figure*}

 \begin{figure*}[t!]
    \centering
    \includegraphics{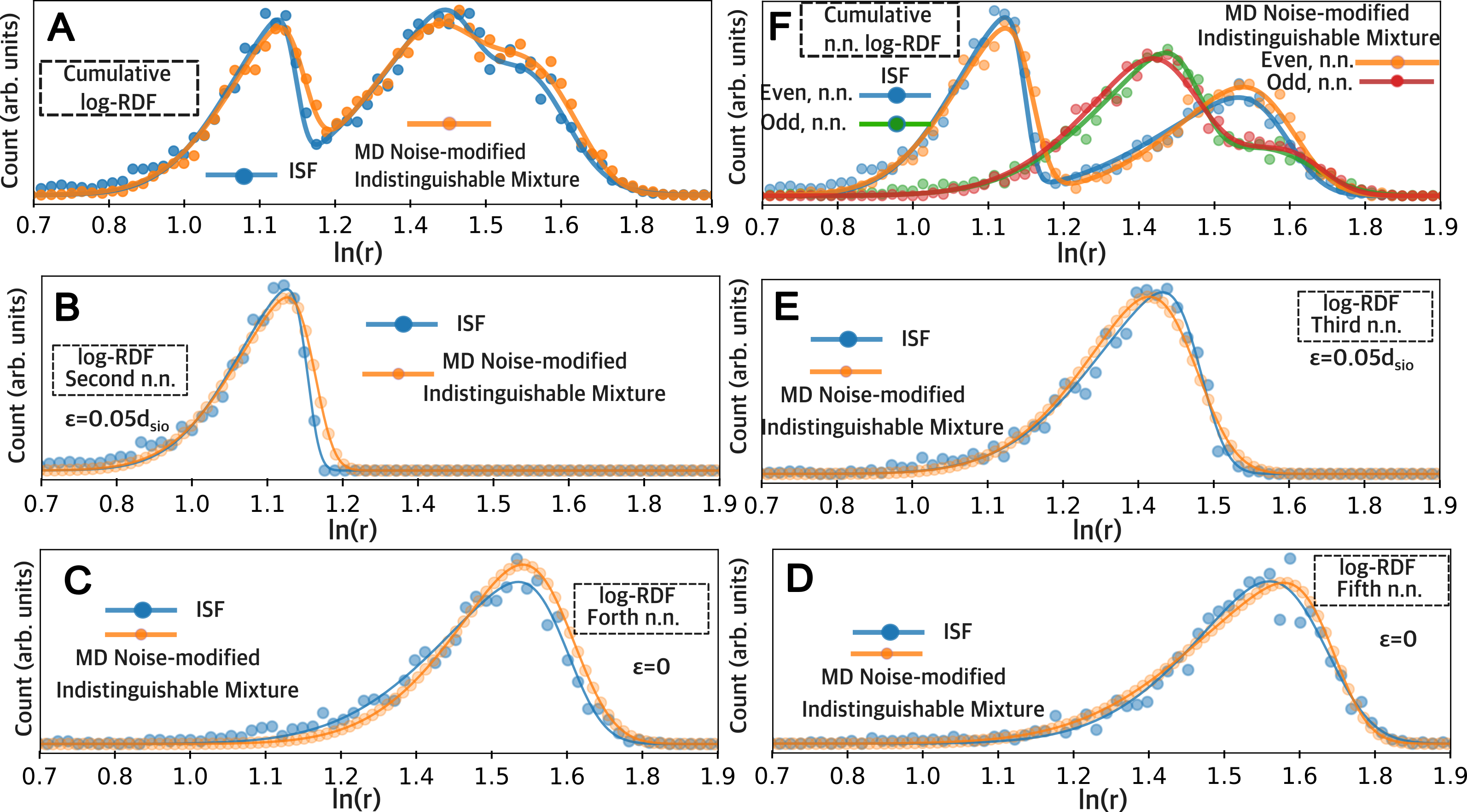}
    \caption{\textbf{A:} Comparison of the log radial distribution function (log-RDF) between ISF N-NRS set and noise-modified MD N-NRS set, reconstructed from the sum of all nearest neighbor log-RDFs, given in \textbf{B-E}. The resulting indistinguishable mixture model (in orange) is detailed in the main text and in Fig.~\ref{fig7}. \textbf{B-E:} individual nearest neighbor log-RDFs contributions, sampled from the noise-modified MD N-NRS set (in orange) and ISF N-NRS set, respectively. \textbf{F:} The sum of odd nearest neighbor log-RDF (in red and green color) and the sum of even nearest neighbor log-RDF (in orange and green color). The red-colored and orange-colored log-RDFs are sampled from the noise-modified MD N-NRS set, and green-colored and blue-colored log-RDFs are sampled from the ISF N-NRS set, respectively (as also indicated in the legend). The ISF log-RDF closely matches the log-RDF obtained from the indistinguishable mixture model. This supports the earlier conclusion that the ISF framework captures the constituent-independent portion of the total structural anisotropy statistics. The statistics for nearest-neighbor distances are calculated using 1000 N-NRSs with \( \theta = 0.005\pi \), \( n = 3 \times 10^4 \), sampled from the 10-NRS set.} 
    \label{fig8}
\end{figure*}

\textbf{Local structural fluctuations and higher order mixture models} In order to approximate constituent-independent contributions to the anisotropy statistics, illustrated in Fig.~\ref{fig7}F, we incorporate local structural fluctuations into each MD-generated N-NRS:
\begin{equation}
    [\mathbf{X}^N] \to [\mathbf{X}^N+\boldsymbol{\epsilon}^N],
\end{equation}
where $\boldsymbol{\epsilon}^N \sim \mathcal{N}(0,\epsilon)$ is a multi-dimensional Gaussian random variable with standard deviation $\epsilon$. We refer to this as the \textit{noise-modified} N-NRS set,
\begin{equation}
\mathcal{E}^{\mathrm{nme}}_N = \{[\mathbf{X}^N_u + \boldsymbol{\epsilon}^N_u], [\mathbf{X}^N_{u'} + \boldsymbol{\epsilon}^N_{u'}], [\mathbf{X}^N_{u''} + \boldsymbol{\epsilon}^N_{u''}], \cdots \}, 
\end{equation}
similar to the set defined in Eq.~\ref{ensemble}. 

The noise-modified log-RDF (shown in orange in Fig.~\ref{fig7}A) demonstrates that adding structural fluctuations to the MD-generated N-NRS set significantly reduces the characteristic peaks in the log-RDF, in contrast to the red curve obtained without noise. Both are compared to the log-RDF of the ISF-generated N-NRS set (shown in blue).

As shown in the nearest-neighbor log-RDFs in Fig.~\ref{fig7}D-E, the odd nearest-neighbor distances uniquely determine the $[\text{Si-O}]$ pair distances. By introducing Gaussian structural fluctuations with variance $\epsilon \sim 0.05d_{\mathrm{iso}}$, as seen in Fig.~\ref{fig8}E, the constituent-dependent contributions are averaged out. The third (odd) noise-modified nearest-neighbor log-RDF (orange) closely aligns with the ISF-derived statistics (blue).

The constituent-dependent contributions to anisotropy statistics diminish with increasing distance, and even without introducing structural fluctuation ($\epsilon = 0$), this trend is reflected in the nearest-neighbor log-RDFs shown in Fig.~\ref{fig8}C-D.

For even nearest neighbors, unlike the odd case, we must simultaneously incorporate contributions from both $[\text{O-O}]$ and $[\text{Si-Si}]$ to establish indistinguishability.

We propose that, to leading order, the constituent-dependent contribution to the MD structural anisotropy statistics can be screened out by considering three sources: $[\text{O-O}]$, $[\text{Si-Si}]$, and $[C] = ([\text{Si-Si}] + [\text{O-O}]) / 2$, with corresponding nearest neighbor log-RDFs, $\Tilde{g}([\text{O-O}])$, $\Tilde{g}([\text{Si-Si}])$, and $\Tilde{g}([C])$, respectively.

The first two are direct contributions, while the third captures local structural correlations that vanish at large distances. The mixture model is drawn from the noise-modified N-NRS set above, and is expressed as (up to normalization):
\begin{equation}
\Tilde{g}(X) \sim \alpha \Tilde{g}([\text{O-O}]) + \beta \Tilde{g}([\text{Si-Si}]) + (1-\alpha-\beta) \Tilde{g}([C]),
\label{mix}
\end{equation}
where $\alpha,\beta$ are mixture constituent-dependent coefficients.

As illustrated in Fig.~\ref{fig7}B-C, the indistinguishability of this mixture can be interpreted as follows: each observation of structural anisotropy arises from randomly selecting either $[\text{O-O}]$ (yellow), $[\text{Si-Si}]$ (purple), or $[C]$ (pink), with probabilities $\alpha$, $\beta$, and $1 - \alpha - \beta$, respectively. The observed structural anisotropy statistics are thus given by the weighted sum of these three distributions (in orange).

\begin{figure*}[h!]
    \centering
    \includegraphics{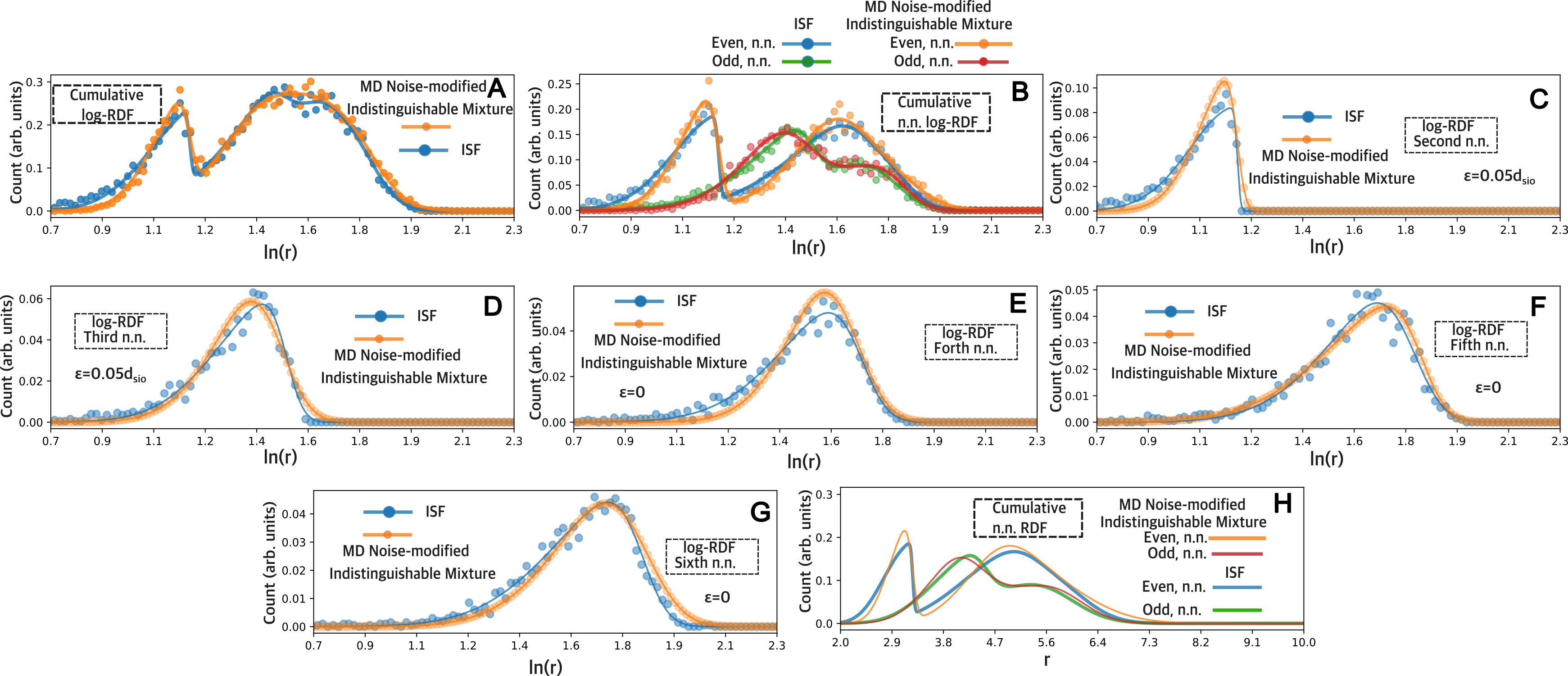}
    \caption{\textbf{A-G:} Nearest neighbor log-RDF contributions for $N=12$, with axis labels and legends identical to those in Fig.~\ref{fig8}.  
\textbf{H:} Partial sum of nearest neighbor contributions to the cumulative log-RDF, generated by ISF and MD simulations. } 
    \label{fig9}
\end{figure*}

\begin{figure*}[t!]
    \centering
    \includegraphics{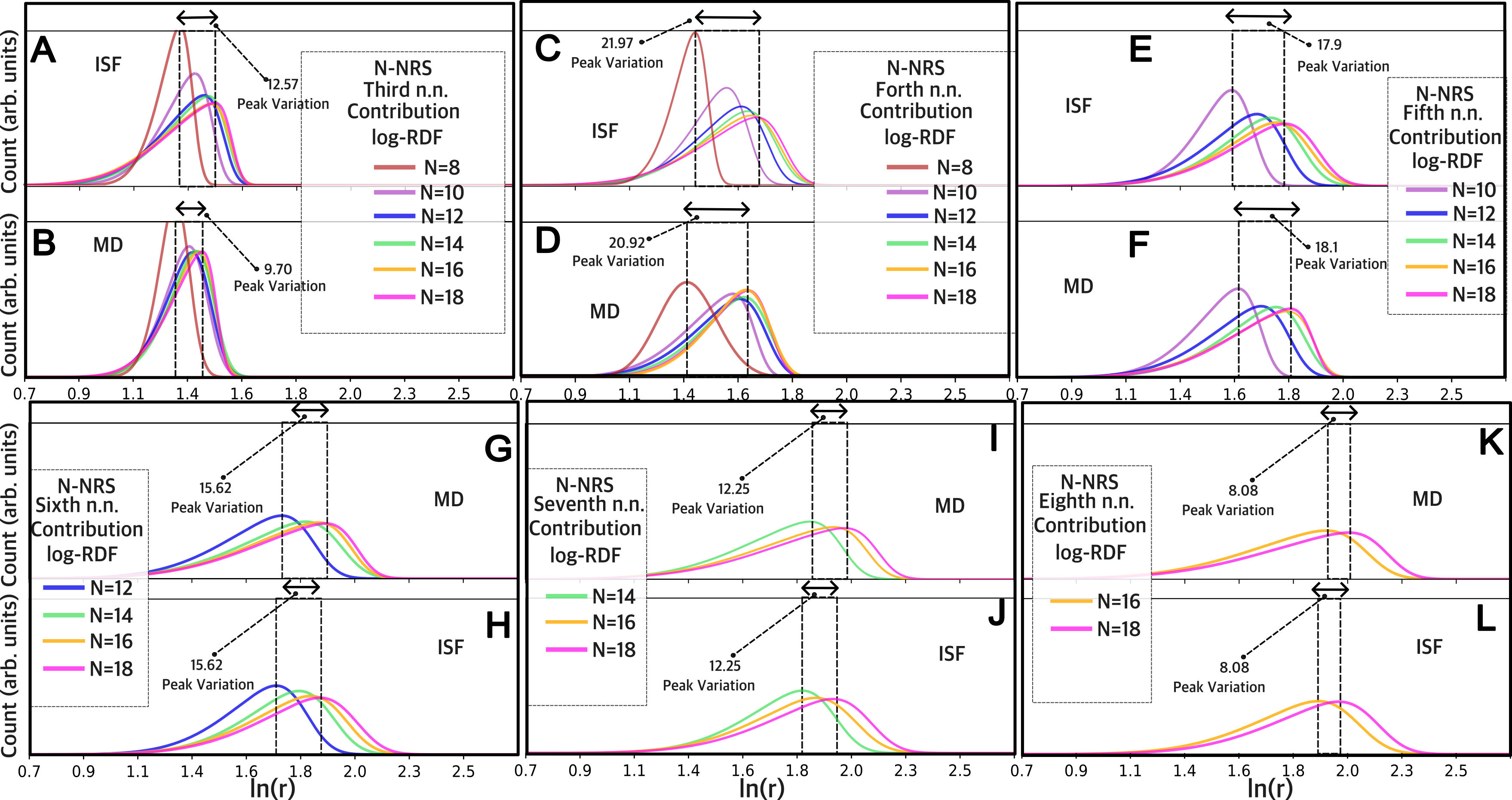}
    \caption{
\textbf{A-L:} Third to eighth nearest neighbor log-RDF contributions, obtained from ISF and MD simulations. Each color represents the log-RDF contribution of a specific N-NRS set, ranging from $N=8$ to $N=18$. The peak variation per N (in pixels with arbitrary units) is indicated in each panel. For example, for the range $8 \to 18$, the peak variation is computed as $\text{peak variation} = \text{value}/5$. Nearest neighbor statistics are computed using 1000 N-NRSs with \( \theta = 0.1\pi \).} 
    \label{fig10}
\end{figure*}

\begin{figure*}[t!]
    \centering
    \includegraphics{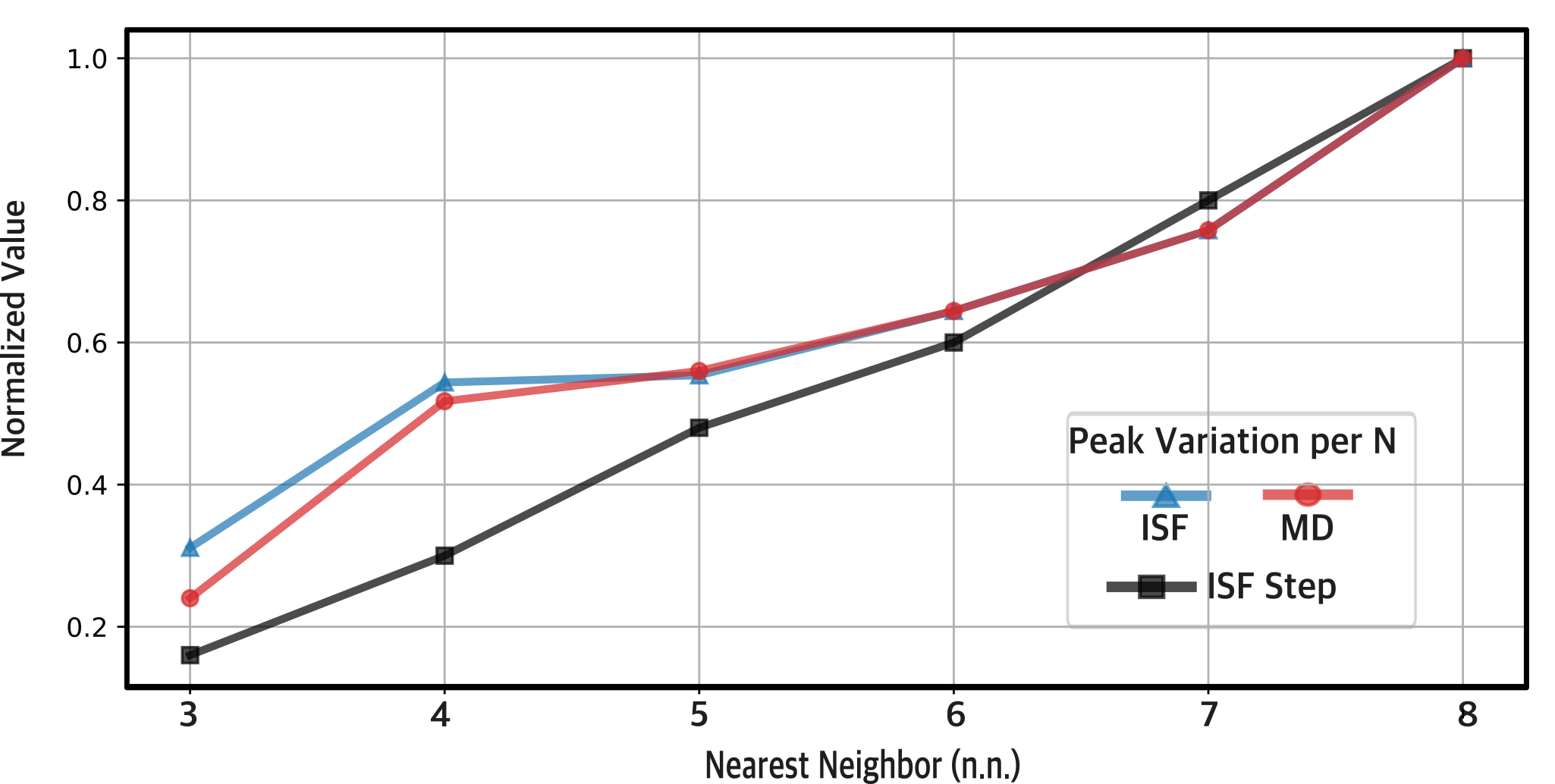}
    \caption{Change in peak variation per N (Fig.~\ref{fig10}) with respect to nearest neighbor contributions, sampled from MD-generated (red) and ISF-generated (blue) N-NRS sets. The Markov steps \( n = 40, 75, 120, 150, 200, 250 \) are normalized and used as a reference (black). The close agreement in peak variation per N supports the conclusion that the ISF formulation captures the constituent-independent contributions produced by MD simulations. } 
    \label{fig11}
\end{figure*}

\begin{figure*}[t!]
    \centering
    \includegraphics{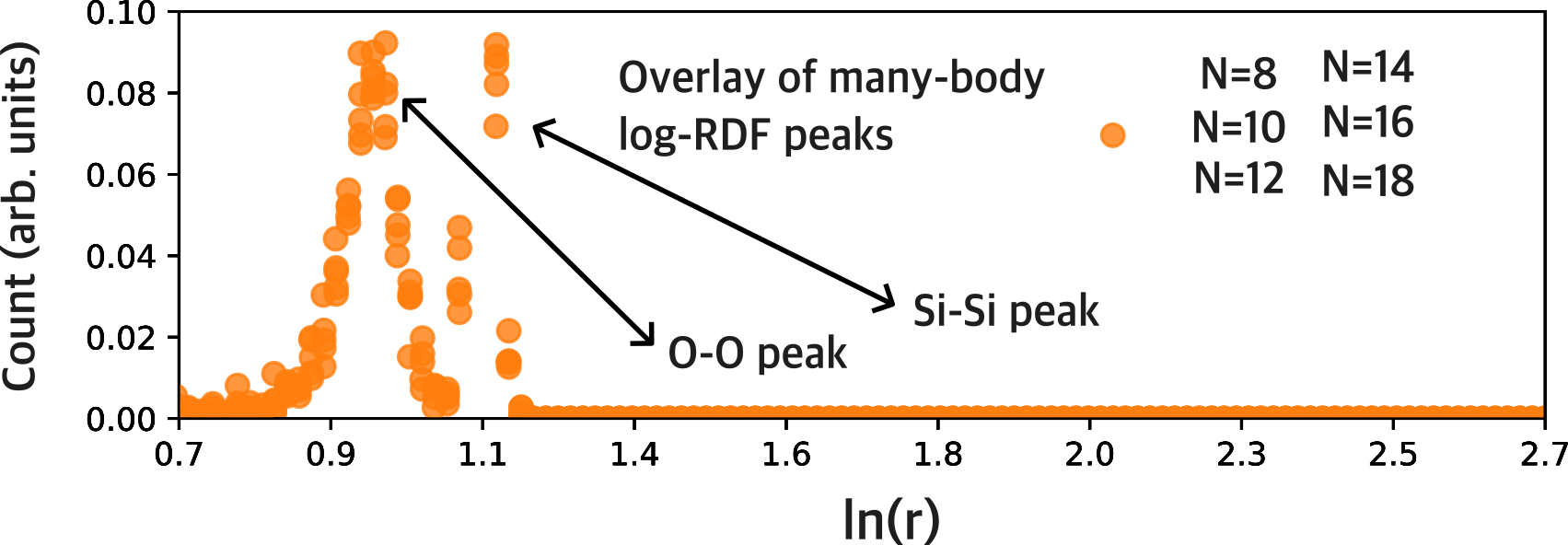}
    \caption{ As discussed in the main text, the many-body peaks \cite{10.1063/1.3632968,Mozzi:a07098} corresponding to oxygen-oxygen ($d\sim  0.27\mathrm{nm}$) and silicon-silicon ($d\sim  0.31\mathrm{nm}$) interactions are dominant, constituent-dependent features. Across the range $N = 8 \to 18$, these peaks remain invariant and are not supposed to be captured by the ISF formulation.} 
    \label{fig12}
\end{figure*}

We numerically reconstruct the N-NRS log-RDF and log-RDFs and demonstrate consistency between the ISF-generated and the constituent-independent portions of the MD-generated statistics, as shown in Fig.~\ref{fig8}A-F. In addition to the log-RDF reconstruction demonstrated for N-NRSs with $N=10$ in Fig.~\ref{fig8}, we present the log-RDF reconstructions for $N=12$ (Fig.~\ref{fig9}), which lead to similar conclusions.

\end{document}